\documentclass[aps,prd,amssymb,eqsecnum,showpacs,nofootinbib,floatfix,twocolumn,a4paper]{revtex4-1}

\usepackage{mathrsfs}
\usepackage{latexsym,amsmath,amsfonts,amssymb}
\usepackage{bm}

\usepackage{graphicx}

\allowdisplaybreaks

\newcommand{\be}{\begin{equation}}
\newcommand{\ee}{\end{equation}}
\newcommand{\bea}{\begin{eqnarray}}
\newcommand{\eea}{\end{eqnarray}}

\newcommand{\md}{\mathrm{d}}

\newcommand{\D}{\partial}
\newcommand{\e}{\varepsilon}
\newcommand{\cG}{{\cal G}}
\newcommand{\cL}{{\cal L}}
\newcommand{\cM}{{\cal M}}

\begin{document}

\title{On the four-loop static contribution to the gravitational interaction potential \\ of two point masses}

\author{Thibault Damour}
\email{damour@ihes.fr}
\affiliation{Institut des Hautes Etudes Scientifiques, 35 route de Chartres, 91440 Bures-sur-Yvette, France}

\author{Piotr Jaranowski}
\email{p.jaranowski@uwb.edu.pl}
\affiliation{Faculty of Physics,
University of Bia{\l}ystok,
Cio{\l}kowskiego 1L, 15--245 Bia{\l}ystok, Poland}

\date{\today}

\begin{abstract}
We compute a subset of  three, velocity-independent four-loop (and fourth post-Newtonian) contributions to the
harmonic-coordinates effective action of a gravitationally interacting system of two point-masses.
We find that, after summing the three terms, the coefficient of the total contribution is rational,
due to a remarkable cancellation between the various occurrences of $\pi^2$. This result, obtained by a 
classical field-theory calculation,  corrects the recent effective-field-theory-based calculation by
Foffa et al.\ [arXiv:1612.00482]. Besides showing the usefulness of the saddle-point approach
to the evaluation of the effective action, and of ${\bm x}$-space computations, our result
brings a further confirmation of the current knowledge of the fourth post-Newtonian effective action.
We also show how the use of the generalized Riesz formula [Phys.\ Rev.\ D {\bf 57}, 7274 (1998)]
allows one to {\it analytically} compute a certain four-loop scalar master integral (represented by a four-spoked wheel
diagram) which was, so far, only numerically computed.
\end{abstract}

\maketitle

\section{Introduction}

The analytical study, to ever-increasing accuracy, of the motion and radiation of two compact bodies
(with comparable masses) in General Relativity has been vigorously pursued over the last decades, 
with the aim of helping the construction of accurate templates for the data-analysis pipeline of the 
network of ground-based interferometric gravitational-wave detectors.  And indeed, the bank of 250 000
templates used in the matched-filter searches and data-analyses of the first observing run of advanced 
LIGO \cite{TheLIGOScientific:2016pea} have been defined \cite{Taracchini:2013rva} within the analytical effective one-body (EOB) formalism 
\cite{Buonanno:1998gg,Buonanno:2000ef,Damour:2000we,Damour:2001tu,Damour:2008gu}. 
The EOB formalism combines, in a suitably resummed format, perturbative, analytical [post-Newtonian (PN)] results on the motion and
radiation of compact binaries, with some non-perturbative information extracted from numerical
simulations of coalescing black-hole binaries.

In this work we focus on the  conservative dynamics of two spinless bodies. The current level of accuracy 
on the analytical knowledge of this problem is the fourth post-Newtonian (4PN) accuracy. The 4PN
Hamiltonian [in Arnowitt-Deser-Misner (ADM) coordinates] of two mass points\footnote{It was shown long ago \cite{Damour1983} that the extension effects of compact bodies
show up only at the 5PN level, so that they can be modelled by point masses below the 5PN accuracy.} is non-local in time, and was first obtained  in complete form in Ref.\  \cite{Damour:2014jta},
based on the computation of the local contributions in Ref.\ \cite{Jaranowski:2015lha}.
(Earlier, partial results were obtained in Refs.\ \cite{Foffa:2012rn,Jaranowski:2012eb,Jaranowski:2013lca,Bini:2013zaa}.)
The non-local action of Ref.\  \cite{Damour:2014jta} was reduced to a local Hamiltonian in Ref.\ \cite{Damour:2015isa}.
(This ``local reduction" was obtained by using an expansion in powers of the eccentricity, together with suitable redefinitions of the phase-space variables, as detailed in \cite{Damour:2016abl}.)
Since then, the only other attempt to derive the complete 4PN dynamics has been the harmonic-coordinates calculation of Ref.\ \cite{Bernard:2015njp}.
Most of the terms in the action of Ref.\ \cite{Bernard:2015njp} agree with the results of Refs.\ \cite{Damour:2014jta,Damour:2015isa}, except a couple of them. 

To discuss the discrepancies between the harmonic-coordinates result of Ref.\ \cite{Bernard:2015njp} and the ADM-coordinates
one of Refs.\ \cite{Damour:2014jta,Damour:2015isa,Jaranowski:2015lha}, it is convenient to order the various contributions to 
the interaction Hamiltonian (which starts by the Newtonian one $-G m_1 m_2/r_{12}$) by means of the powers of the symmetric 
mass ratio $\nu$. Our notation (besides using $G$ for Newton's gravitational constant) is 
\be
 M \equiv m_1 +m_2;\:
 \mu \equiv \frac{m_1 m_2}{m_1+m_2};\:
 \nu  \equiv \frac{\mu}{M} = \frac{m_1 m_2}{(m_1+m_2)^2}.
\ee
We denote the  two masses of the binary system as $m_1$ and $m_2$, while $r_{12} = |{\bm r}_{12}|$ 
(where ${\bm r}_{12}\equiv  {\bm x}_1 - {\bm x}_2$)
denotes the relative distance. We work here in the center-of-mass system; when doing so in a Hamiltonian framework, one considers
the ratio ${\bm p}/\mu$ (where ${\bm p} ={\bm p}_1= - {\bm p}_2$) as fixed.
The $\mu$-reduced Hamiltonian
${\hat H}_{\text{4PN}} \equiv {\hat H}_{\text{4PN}}/\mu$ can then be decomposed as
\begin{align}
\label{nuexp4pn}
{\hat H}_{\text{4PN}}\left(\frac{{\bm r}_{12}}{GM}, \frac{ {\bm p}}{\mu}\right) 
&= {\hat H}^{\text{4PN}}_0+ \nu {\hat H}^{\text{4PN}}_1+ \nu^2 {\hat H}^{\text{4PN}}_2
\nonumber \\
&\quad + \nu^3 {\hat H}^{\text{4PN}}_3+ \nu^4 {\hat H}^{\text{4PN}}_4.
\end{align}
Here, ${\hat H}^{\text{4PN}}_0$ describes the 4PN-level contribution to the dynamics of a test mass moving around
a central body of mass $M=m_1+m_2$, while $\nu {\hat H}^{\text{4PN}}_1$ describes the first self-force (1SF) correction to
the latter test-mass dynamics,  $ \nu^2  {\hat H}^{\text{4PN}}_2$ the second self-force (2SF) correction, etc. [In diagrammatic
language, computing 1SF effects on the ``small mass" $m_1$ (say) corresponds to computing one gravitational loop in the external 
gravitational field of a black hole of mass $m_2 \gg m_1$.]

It was shown in Ref.\ \cite{Bernard:2015njp} that all the terms that are non-linear in $\nu$
[i.e.\ $\nu^2 {\hat H}^{\text{4PN}}_2 + \nu^3 {\hat H}^{\text{4PN}}_3+ \nu^4 {\hat H}^{\text{4PN}}_4$ in Eq.\ \eqref{nuexp4pn}]
in their harmonic-coordinate result agree (modulo a suitable contact transformation) with the ADM action
of Ref.\ \cite{Damour:2014jta}. The discrepancies are limited to the $\nu$-linear (1SF-level)
contribution $  \nu {\hat H}^{\text{4PN}}_1$. It was later shown in Ref.\ \cite{Damour:2016abl} that  the $\nu$-linear terms in 
the local reduction \cite{Damour:2015isa} of the ADM non-local action were  in full agreement with several different
(analytical and numerical) gravitational self-force computations (combined with results from EOB theory, and from the first law
of binary dynamics  \cite{LeTiec:2011ab,Blanchet:2012at,Tiec:2015cxa}), and it was concluded that several claims, and results, of
Ref.\ \cite{Bernard:2015njp} were incorrect, and must be corrected both by evaluating the energy in keeping with 
Refs.\ \cite{Damour:2014jta,Damour:2015isa}, and by the addition of a couple of {\it ambiguity parameters} linked
to subtleties in the regularization of infrared and ultraviolet divergences. The values of the needed additional ambiguity
parameters (denoted there $\Delta a$ and  $\Delta b$, when using the ``gauge" $c=0$) were determined in \cite{Damour:2016abl}
to be $\Delta a^{\rm tot}=\Delta a - 11\,\frac{16}{15}\,\Delta C= \frac{2179}{315}$ and
$\Delta b = + 12\,\frac{16}{15}\,\Delta C= - \frac{192}{35}$ [inserting Eqs.\ (6.1), (7.4) of \cite{Damour:2016abl} in
Eq.\ (6.3) there]. Recently, Ref.\ \cite{Bernard:2016wrg} confirmed all those conclusions, and notably the values of the
ambiguity parameters (which they denote $-\delta_1 \equiv \Delta a^{\rm tot}$ and $- \delta_2\equiv \Delta b$)
that must be added to the harmonic-coordinates Hamiltonian to correct it.

Very recently, Ref.\ \cite{Foffa:2016rgu} applied the so-called effective field theory (EFT) method \cite{Goldberger:2004jt} to
the computation of a {\it subset} of the contributions to the harmonic-coordinates Lagrangian $L$. Given some specified {\it gauge-fixing}
additional contribution to the Einstein-Hilbert action [here the standard harmonic-coordinates gauge-fixing term 
$S_{gf}=(16 \pi G)^{-1} \int \md^Dx \sqrt{g}(-\frac12 g_{\mu \nu} \Gamma^{\mu} \Gamma^{\nu})$ with 
$\Gamma^{\mu} \equiv g^{\rho \sigma} \Gamma^{\mu}_{\rho \sigma}$], the effective\footnote{The reduced action 
(obtained by ``integrating out" the mediating field) describing the conservative dynamics of some particles is called by various names:
Fokker action, reduced action, effective action, \ldots\,. Here, we shall use the name ``effective action" to avoid confusion with the
``order-reduced" local action \cite{Damour:2015isa} which replaces the original non-local-in-time 4PN action by an equivalent local-in-time one.}
action, $S_{\rm eff} = \int \md t L$, describing the conservative dynamics of the binary system can be decomposed in powers
of $G$ and of the velocities ${\bm v}_a$, $a=1,2$ (together with their various time derivatives $\dot{\bm v}_a$, $\ddot{\bm v}_a$, $\cdots$)\footnote{Here, we formally consider
the non-local-in-time piece of the (interaction) action as a functional of the infinite set of time derivatives of ${\bm v}_a$.}.
In particular, the structure of the  interaction Lagrangian (say up to the 4PN level) is roughly described by expanding the $n$-th power
(with $n\leq 4$) on the first rhs of the following sketchy formula (where $c$ denotes the velocity of light):
\begin{align}
\label{Gexp4pn}
&L_{\leq {\text{4PN}}}^{\rm int}[{\bm x}_a, {\bm v}_a, \dot{\bm v}_a, \cdots]
\nonumber \\
&\qquad \sim \frac{G m_1 m_2}{r_{12}} \sum_{n \leq 4}\left(  \frac{Gm}{r c^2} + \frac{v^2}{c^2} + \frac{r_{12} \dot v}{c^2} + \cdots  \right)^n 
\nonumber \\
&\qquad \sim \frac{G m_1 m_2}{r_{12}} \sum \left( \frac{Gm}{r c^2} \right)^{n_1}  \left( \frac{v^2}{c^2} \right)^{n_2} \left( \frac{r_{12} \dot v}{c^2} \right)^{n_3} \cdots\,.
\end{align}
In the multiple sum on the last rhs the sum of the powers $n=n_1+n_2+n_3 +\cdots$ must be $\leq 4$.
As will be described in more detail below, the various contributions in the fully expanded form of $ L_{\leq {\textrm{4PN}}}^{\rm int}$ can
be described in terms of Feynman diagrams. Here, following Ref.\ \cite{Foffa:2016rgu}, we shall focus on the contributions
having the highest possible power of $G$, i.e.\ $n_1=4$, $0=n_2=n_3=\cdots$ in Eq.\ \eqref{Gexp4pn}, corresponding
to a purely ``static" term, {\it quintic} in $G$, without effects linked to velocities, or derivatives of velocities.

It was understood long ago \cite{Damour:1985mt,Damour:1990jh} that any term that is
non-linear in the derivatives of velocities can be eliminated from a higher-order Lagrangian $L(x, v, \dot v, \cdots)$
by adding suitable ``double-zero" terms [quadratic in $\dot v- (\dot v)^{\text{on-shell}}$], thereby allowing
one to replace a general higher-order Lagrangian by an equivalent simpler one that is {\it linear} in accelerations.
(A further reduction, involving a redefinition of the particle variables allows one to eliminate the
accelerations \cite{Schafer:1984mr,Damour:1985mt,Damour:1990jh}.) The procedure of reduction of terms
quadratic (or more) in accelerations to a linear dependence in accelerations involves the on-shell
equations of motion $(\dot v)^{\text{on-shell}} \sim G m r_{12}^{-2} (1 + O(1/c^2))$, and thereby introduces
a mixing between the various powers of $G$ in the expanded Lagrangian Eq.\ \eqref{Gexp4pn}.
In particular, after reduction to a $\dot v$-linear form (as was done in \cite{Bernard:2015njp}), the contribution
proportional to $G^5$ is given by a sum of terms coming from terms $\sim G^{1+n}$ in Eq.\ \eqref{Gexp4pn}
having $n\leq 4$. More precisely, as terms quadratic in accelerations contain at least two powers of $1/c^2$, we have
\be
\label{newGexp4pn}
 L_{{\text{4PN}}}^{\rm int}\big|^{O(G^5)}_{{\rm linear} \ {\rm in}\ \dot{\bm v}_a}
 \subseteq L_{\leq {\text{4PN}}}^{\rm int}\big|^{(n=4)} + \sum_{n=0,1,2} L_{\leq {\text{4PN}}}^{\rm int}\big|^{(n)}_{\dot v^2},
\ee
with values $n=0,1,2$.  

Foffa et al.\  pointed out \cite{Foffa:2016rgu} three facts: (i) the terms non-linear in accelerations coming
from $n=0$ and $n=1$ on the rhs of Eq.\ \eqref{newGexp4pn} only contribute {\it rational} coefficients to the lhs;  (ii)
the terms quadratic in accelerations coming from  $n=2$ on the rhs contribute the following $\pi^2$-dependent 4PN 
$O(G^5)$ terms to $L_{{\text{4PN}}}^{\rm int}\big|^{O(G^5)}_{{\rm linear} \ {\rm in}\ \dot{\bm v}_a}$
\be \label{pi2froma2}
\frac{105}{32} \pi^2 \frac{G^5 (m_1^4 m_2^2+ m_1^2 m_2^4)}{c^8 r_{12}^5} - \frac{71}{16} \pi^2 \frac{G^5 m_1^3 m_2^3}{c^8 r_{12}^5};
\ee
and, (iii) the $\pi^2$-dependent terms \eqref{pi2froma2} [coming from the $O(G^3 \dot v^2$) action] coincide with the  
($v$- and $\dot v$-independent) $\pi^2$-dependent terms present in the full, linear-in-acceleration harmonic-coordinates 4PN Lagrangian
derived in Ref.\ \cite{Bernard:2015njp} [see Eq.\ (5.6f) there]. 

As the latter contributions in the harmonic-coordinates Lagrangian of \cite{Bernard:2015njp} agree with
corresponding contributions in the ADM Hamiltonian of \cite{Damour:2014jta}, one would then conclude (barring
a coincidental agreement between two incorrect results) from
Eq.\ \eqref{newGexp4pn} that the coefficients entering the $n=4$ [i.e.\ $O(G^5)$] contribution to the original (non-linear in derivatives of $v$) 
4PN effective Lagrangian $L_{\leq {\text{4PN}}}^{\rm int}\big|^{(n=4)}$ should not contain any $\pi^2$, i.e.\ should be a rational number.
In other words, there should be no new, genuine $\pi^2$ at the  $O(G^5)$ level. 

However, Foffa et al.\ \cite{Foffa:2016rgu} have recently reported the computation, within the EFT approach,
of the 50 Feynman diagrams contributing to the $n=4$ [i.e.\ $O(G^5)$] contribution
to $L_{\leq {\text{4PN}}}^{\rm int}[{\bm x}_a,{\bm v}_a,\dot{\bm v}_a,\cdots]$ in Eq.\ \eqref{Gexp4pn}.
Their results comprise three contributions with $\pi^2$-dependent coefficients, namely 
\begin{subequations}
\label{fmss}
\begin{align}
L_{33}^{\text{FMSS}} &=  (32 - 2 \pi^2)  \frac{G^5 m_1^3 m_2^3}{c^8 r_{12}^5} ,
\\[1ex]
L_{49}^{\text{FMSS}} &= (64 - 6 \pi^2) \frac{G^5 m_1^3 m_2^3}{c^8 r_{12}^5} , 
\\[1ex]
L_{50}^{\text{FMSS}} &= \left(\frac{248}{9} - \frac{8}{3} \pi^2\right) \frac{G^5 m_1^3 m_2^3}{c^8 r_{12}^5} .
\end{align}
\end{subequations}
Note that we cited here {\it twice} the quantitities respectively denoted ${\cal L}_{33}$, ${\cal L}_{49}$ and ${\cal L}_{50}$
in \cite{Foffa:2016rgu} because it seems that they implicitly assume that the $m_1$-$m_2$ symmetric Lagrangian contributions
${\cal L}_{33} \sim {\cal L}_{49} \sim {\cal L}_{50} \sim m_1^3 m_2^3 $ should be augmented by their $ 1 \leftrightarrow 2$
images, and thereby doubled.

The sum of the three contributions \eqref{fmss} contains the $\pi^2$-dependent term
\be
- \frac{32}{3} \pi^2 \frac{G^5 m_1^3 m_2^3}{c^8 r_{12}^5}
\ee
which disagrees with the result of \cite{Bernard:2015njp} (which is derived with the use of the same, harmonic gauge-fixing term). 
In terms of the $\mu$-reduced Hamiltonian \eqref{nuexp4pn},
this discrepancy is proportional to $\nu^2$, and therefore at the 2SF level. All the contributions $O(\nu^2)$ to the $\mu$-reduced
4PN action agreed (modulo a contact transformation) between the two existing complete 4PN calculations \cite{Damour:2014jta} and \cite{Bernard:2015njp}.

The main aim of the present paper is to perform a new, independent calculation of the three contentious Lagrangian
contributions ${ L}_{33}$, ${ L}_{49}$ and ${ L}_{50}$ to decide whether there were subtle, hidden errors in \cite{Bernard:2015njp} 
and \cite{Damour:2014jta} that coincidentally agree, or whether there is an error in the EFT-theory evaluation of the corresponding
Feynman integrals. A secondary aim of the present paper concerns the analytical computation of a certain $d$-dimensional, four-loop 
``master" Feynman integral, denoted ${\cal M}_{3,6}$ in Ref.\ \cite{Foffa:2016rgu}. This master integral contributes to the
values of both ${ L}_{33}$, and ${ L}_{50}$. Though they employed some of the most advanced Feynman-integral computation
techniques, Foffa et al. did not succeed in analytically evaluating the $d$-dimensional, four-loop integral ${\cal M}_{3,6}$, and had to resort to
a many-digit numerical evaluation of the coefficients of the Laurent expansion of ${\cal M}_{3,6}(d)$ in powers of $ \e \equiv d-3$.
This  evaluation gave very solid numerical evidence for the presence of $\pi^2$ at the $\e^0$ level, and this has been assumed
to be exactly true in the computation of the results Eqs.\ \eqref{fmss}.

The two main results of the present paper will be: (i) to show that one can {\it analytically} evaluate (by notably using the
generalized Riesz formula derived in \cite{Jaranowski:1997ky}, which was also crucial to the computation of the local ADM 4PN 
Hamiltonian computation \cite{Jaranowski:2015lha}) the relevant first three
terms in the $\e$ expansion of the four-loop master integral ${\cal M}_{3,6}(d=3+ \e)$, and, in particular, rigorously prove the presence
of $\pi^2$ at the $\e^0$ level ; and (ii) explain away the seeming contradiction following from the presence 
of $\pi^2$ in the EFT evaluation  \eqref{fmss} of the four-loop
integrals  ${ L}_{33}$, ${ L}_{49}$ and ${ L}_{50}$, by showing that a new, independent calculation of these integrals
(using, instead of the EFT technique of \cite{Foffa:2016rgu}, the  alternative,
diagrammatic ``field theory" approach to the effective action introduced long ago by 
Damour and Esposito-Far\`ese \cite{Damour:1995kt}, together with  ${\bm x}$-space
techniques, and the use of the generalized Riesz formula), leads to results that crucially
differ from the ones cited above in that $\pi^2$ simply {\it cancels out} in the sum  ${ L}_{33} +{ L}_{49} +{ L}_{50}$.

\section{Various approaches to the effective action for gravitationally interacting point masses}

The introduction of a  classical ``variational principle that takes account of the mutual interaction of multiple particles
without introducing fields" dates back to Fokker's 1929 definition \cite{Fokker1929} of the following relativistic functional of several worldlines
(labelled by $a,b=1,\ldots,N$) describing $N$ electromagnetically interacting charged point masses
(here we use $c=1$, and all the quantitites are defined in a Minkowski spacetime of signature mostly plus)
\begin{align}
\label{fokkerem}
S_{\rm eff}^{\rm class}[x_a(s_a)] &= -\sum_a m_a \int \md s_a 
\nonumber \\
& +  \frac{1}{2} \sum_{a,b} e_a e_b \iint \md x_a^{\mu}\, \md x_{b \mu}\, \delta\left(  (x_a-x_b)^2 \right).
\end{align}
The action \eqref{fokkerem} is obtained by classically ``integrating out" the electromagnetic field $A_{\mu}(x)$ in the usual
total relativistic action for the particles and the field, i.e.\ by replacing the (time-symmetric, Lorenz-gauge) solution of the equation of motion of 
$A_{\mu}(x)$ in  presence of given worldlines (say 
$A_{\mu}^{\rm Lorenz}[x ; x_a(s_a)] $ ) in the original particle $+$ field action.

The action \eqref{fokkerem} played a central role in the 1949 work of Wheeler and Feynman \cite{Wheeler:1949hn}.
Let us also note that Fokker's original paper features spacetime diagrams of worldlines interacting via time-symmetric propagators.
It is therefore probable that the introduction of quantum interaction diagrams (or Feynman diagrams) by Feynman around the same time
was partly motivated by Fokker's classical interaction diagrams.
Clear evidence for this is the 1950 paper of Feynman \cite{Feynman:1950ir} in which he introduces the (complex) quantum effective 
action for charged particles
defined (in modern notation) through taking the {\it logarithm} of a functional integral over the field
(in presence of given classical charged worldlines)
\be \label{feynmanem}
e^{ \frac{i}{\hbar} S_{\rm eff}^{\rm quant}}= \int D A_{\mu} e^{ \frac{i}{\hbar}(S_{\rm particle} + S_{\rm field})} .
\ee
He then explicitly shows that $S_{\rm eff}^{\rm quant}$, \eqref{feynmanem}, only differs from its classical counterpart, \eqref{fokkerem},
by the replacement of the (real)  time-symmetric propagator $\delta( (x_a-x_b)^2)$ by the (complex) (St\"uckelberg-)Feynman 
propagator $\delta_+( (x_a-x_b)^2)$, with $\delta_+(x) = \frac{i}{\pi (x^2+ i 0)}= \delta(x^2) + PP \frac{i}{\pi x^2}$.

The gravitational analog of the above classical,  effective action for the general relativistic interaction
of point masses reads \cite{InfeldPlebanski}, 
\begin{align}
\label{fokkerg}
S_{\rm eff}^{\rm class}[x_a(s_a)] =
\left[ S_{\rm pm } + S_{\rm EH} + S_{\rm gf}\right]_{g_{\mu \nu}(x)\to g_{\mu \nu}^{\rm gf}[x_a(s_a)]} ,
\end{align}
where $S_{\rm pm }= - \sum_a \int m_a \sqrt{-g_{\mu \nu}(x_a)\,\md x_a^{\mu}\,\md x_a^{\nu}}$ denotes the point-mass action,
$S_{\rm EH}$ the Einstein-Hilbert action, and $S_{\rm gf}$ a gauge-fixing term, and where $g_{\mu \nu}^{\rm gf}[x_a(s_a)]$
denotes the gauge-fixed solution of Einstein's equations in presence of given worldlines.
 The gravitational analog of the above (formally) quantum, effective action reads\footnote{Note that this definition is misprinted
 in Refs.\ \cite{Foffa:2011ub,Foffa:2016rgu}, where the lhs of Eq.\ \eqref{feynmang} is simply written as $\frac{1}{\hbar} S_{\rm eff}^{\rm quant}$,
 without the exponential, and without the imaginary unit; these omissions being later corrected by considering connected
 diagrams and by multiplying the rhs by $-i$.}
 \be
 \label{feynmang}
e^{ \frac{i}{\hbar} S_{\rm eff}^{\rm quant}}= \int D g_{\mu \nu}\, e^{ \frac{i}{\hbar}( S_{\rm pm } + S_{\rm EH} + S_{\rm gf})} .
\ee
In the classical limit, one can evaluate the (formal) path integral \eqref{feynmang} by the saddle point (or stationary phase)
approximation. As the extrema with respect to  $g_{\mu \nu}(x)$ of the exponent in \eqref{feynmang} are simply classical
solutions of the gauge-fixed Einstein equations, one immediately sees that (formally)
\be
S_{\rm eff}^{\rm quant}[x_a(s_a)] = S_{\rm eff}^{\rm class}[x_a(s_a)] + O(\hbar).
\ee
We recalled the above rather well-known facts to clarify that the so-called EFT method [formally based on \eqref{feynmang}] computes
(in the classical limit, and when considering the conservative\footnote{However, as discussed in \cite{Goldberger:2004jt}
and several subsequent papers, the imaginary part of $S_{\rm eff}^{\rm quant}[x_a(s_a)] $ gives useful information about radiation-damping effects.} dynamics) exactly the same quantity
as the classical, Fokker (or, for that matter) ADM, reduction method \eqref{fokkerg}.

However, the two different definitions of the effective action suggest different technical methods for computing it, and this is where
there is a real practical difference in the traditional PN (or post-Minkowskian) computations of $S_{\rm eff}$, and in the EFT-inspired one.
First, let us recall that long before the EFT method was set up \cite{Goldberger:2004jt}, an  alternative,
diagrammatic ``field theory" approach to the (classically defined) effective action was introduced in Ref.~\cite{Damour:1995kt}.
It was explicitly shown in  \cite{Damour:1995kt} how the pertubative, post-Minkowskian way of solving the gauge-fixed Einstein's equations
(say in harmonic gauge) leads to a (classical, Feynman-like) diagrammatic expansion of the effective action for the
particles of the form (with the normalizations chosen there)
\begin{align}
\label{PMexpSeff}
S_{\rm eff}&= S_{\rm free}+ \left[\frac12  I  \right]_{G m^2}+  \left[ \frac12 V + \frac13 T  \right]_{G^2 m^3}
\nonumber \\
&\quad +  \left[ \frac13 \epsilon + \frac12 Z + F + \frac12 H + \frac14 X  \right]_{G^3 m^4} + \cdots.
\end{align}
Here, each letter $I, V, T, \cdots$ is chosen to evoque a correspondingly shaped diagram, when representing the source
by, say, $\circ$. For instance, the diagram $I$ denotes the vertical concatenation of two sources, $\circ$ and $\circ$, located
at the end points of the $I$, via an intermediate
(time-symmetric) gravitational propagator $\arrowvert$, say $\arrowvert_{\! \! \circ}^{\! \! \circ}$.
(Here, the propagator is defined as minus the inverse of the kinetic term, see more discussion of this choice below.)
In other words, $I$ denotes the one-graviton-exchange diagram of the gravitationally interacting source.
[When decomposing the material, two-body source according to the masses, say $\circ=m_1 \circ_1+\,m_2 \, \circ_2$,
the $I$ diagram gives three contributions: two self-gravity ones, $O(G m_1^2)$ and $O(G m_2^2)$,
and a relativistic Newtonian interaction one: $G m_1 m_2 \, \arrowvert_{{\! \! \circ}_1}^{{\! \! \circ}_2}$.]
Similarly, $T$ denotes a diagram where three sources (located at the end points of the $T$)
are connected via three gravitational propagators that meet at a cubic vertex in the middle of the upper branch of the $T$.
In addition,  Ref.\ \cite{Damour:1995kt} gave explicit rules for computing the numerical coefficients to be put in front of
each diagram to correctly evaluate the effective action\footnote{The explicit coefficients shown in Eq.\ \eqref{PMexpSeff}
above follow from the specific $S_n[\varphi] = \frac1n V_n[\varphi^n]$ vertex normalization chosen in \cite{Damour:1995kt}. 
When absorbing the conventional prefactor $\frac1n$ in the definition of the vertex $V_n$ many of the factors in
the effective action \eqref{PMexpSeff} become unity, and the remaining ones are usual symmetry factors.}.
It is sometimes convenient (to better exhibit the physics contained in the effective action) to draw each individual
source $m_1 \circ_1$ or $ m_2 \circ_2$ as a spacetime worldline. Then each diagram in the post-Minkowskian expansion
\eqref{PMexpSeff} becomes made of concatenated propagators, with some propagators starting on the worldlines,
and intermediate propagators joining either a gravitational vertex, or a worldline.  We shall later give explicit examples of such
spacetime representations of effective-action diagrams, which generalize the representation used by Fokker himself back in 1929.

When further taking the PN expansion of the time-symmetric (scalar)
propagator, say 
\begin{align}
\label{pnexpandedpropagator}
&{\mathcal G}(t,\mathbf{x};t',\mathbf{x}') \equiv -4\pi \left( \Delta - \frac1{c^2} \, \partial_t^2 \right)^{-1}
\nonumber\\
&= -4\pi \left( \Delta^{-1} + \frac1{c^2}\Delta^{-2}\partial_t^2 + \frac1{c^4}\Delta^{-3}\partial_t^4 + \ldots \right) \delta (t-t'),
\end{align}
each post-Minkowskian diagram in the expansion \eqref{PMexpSeff} will generate a sequence of PN-type diagrams (involving
inverse powers of the Laplacian, together with time-derivatives, as propagators). These are now three-dimensional (or $d$-dimensional)
diagrams made of PN-propagators $\Delta^{-n}$ connecting the two point masses $m_1 \delta({\bm x}-{\bm x}_1)$
and $m_2 \delta({\bm x}-{\bm x}_2)$ via some intermediate field points that are integrated over.  It has been known for a long time that
the computation of the effective action at the $n$PN level involves diagrams whose topology features $\leq n$ loops. The topological loops
can be recognized either on the spacetime diagrams, or on the projections as $d$-dimensional diagrams. For instance,  
Fig.\ 1 in Ref.\ \cite{Damour:2001bu} represents a spatial, two-point, three-loop diagram representing a 3PN-level contribution $O(G^4 m_1^3 m_2^3)$ to the effective ADM action of two point masses.
Below, we shall give examples (with four loops, at the 4PN level)
of such spatial diagrams.

Summarizing:  the usual, Fokker-like computation of the PN-expanded gravitational action 
(using either harmonic coordinates, or ADM coordinates,
and using either traditional methods or the field-theory-diagrammatic technique of \cite{Damour:1995kt}) leads to a sum of 
${\bm x}$-space integrals involving the concatenation of PN-propagators $\Delta^{-n}\partial_t^{n+1} \delta (t-t') $
and their joining at intermediate spatial points, with vertices involving two derivatives (because of the
structure of the gravitational action \mbox{$\D \D h h + \D \D h h h + \cdots$}). 

The main points we wanted to emphasize here about
the traditional Fokker-like computation of the effective action are:
(i) all the contributions ot the effective action are explicitly real; (ii) all the integrals are in ${\bm x}$-space; (iii) all the integrations by parts used to reduce integrals to some  ``master" integrals
are done in ${\bm x}$-space; (iv) at each stage of the calculation one keeps track of the numerical coefficients multiplying
each integral, because they are directly furnished by the replacement $g_{\mu \nu}(x)\to g_{\mu \nu}^{\rm gf}[x_a(s_a)]$
of the gauge-fixed solution in (essentially) the Einstein-Hilbert Lagrangian (be it in harmonic guise, or in the ADM one).

By contrast, the EFT approach to the effective action is based on expanding  functional integrals of the type (here written,
for pedagogical purposes, as a  scalar toy-model, with a source $s(x)$, taken simply as a linear coupling here), 
 \be \label{feynmanscalar}
e^{ \frac{i}{\hbar} S_{\rm eff} }= \int D \varphi\, e^{ \frac{i}{\hbar}( \int [\frac12 \varphi {\cal K } \varphi + \varphi s  + g \varphi^3 + \cdots]} .
\ee
Instead of expanding around the saddle point of the exponent (as done in the usual Fokker approach) one expands the functional
integral around the Gaussian approximation defined by the free term with kinetic operator ${\cal K }$, and 
with elementary contraction given by
\be \label{phiphi}
\langle \varphi(x) \varphi(y) \rangle= \int D\varphi\, e^{ \frac{i}{\hbar} \int [\frac12 \varphi {\cal K } \varphi]} \varphi(x) \varphi(y) =i \hbar {\cal K }^{-1}_{x,y},
\ee
where ${\cal K }^{-1}_{x,y}$ denotes the inverse of the kinetic operator  (i.e.\ a  Green's function).
One then expands the exponent on the rhs of \eqref{feynmanscalar},
\be
\int D \varphi\,e^{\frac{i}{\hbar} \int [\frac12 \varphi {\cal K } \varphi]}
\sum_n \frac{(i/\hbar)^n}{ n!} \left(\int(\varphi s  + g \varphi^3 + \cdots)\right)^n
\ee
applying Wick's theorem to compute all the $\varphi$ contractions arising from the various powers $ i^n\left(\int(\varphi s  + g \varphi^3 + \cdots)\right)^n/n!$
coming from the expansion of the exponential. (We henceforth set $\hbar=1$ for simplicity.)
The lowest-order contribution comes from the term quadratic in $s$,
namely $\frac{i^2}{2}\langle \int\md x\, \varphi(x) s(x) \int\md y\, \varphi(y) s(y) \rangle = \frac{ i^3}{2} \iint \md x \md y\,  {\cal K }^{-1}_{x,y} s(x) s(y)$.
Factoring one power of $i$ this contributes $\frac{ i^2}{2} \iint \md x \md y\,  {\cal K }^{-1}_{x,y} s(x) s(y) = - \frac12 \iint \md x \md y\,  {\cal K }^{-1}_{x,y} s(x) s(y)$,
to  the effective action $ S_{\rm eff}$.
This indeed coincides with the (correctly normalized) one-quantum exchange energy denoted $+\frac12 I$ above. 

Summarizing: the quantum, Feynman-like computation of the PN-expanded gravitational action deals with a sum of
Wick contractions from the powers of the interaction terms $\varphi s  + g \varphi^3 + \cdots$ in the original field + particle action.
This calculation involves many imaginary units $i$. Because of a certain quantum tradition, these calculations have
been done in $\bm p$-space, rather than in $\bm x$-space, using, e.g., elementary field contractions $ \langle \varphi \varphi \rangle =i/(-p^2)$ if the kinetic term is $\Box$. 
(We use the mostly plus signature.) In this approach one has to take care of correctly
multiplying each diagram by the needed symmetry factor (which can be somewhat tricky when considering high-order contractions).
In doing the explicit calculations at the $n$th PN order, there appear diagrams having up to $n$-loops, corresponding to 
 integrating over $n$ independent loop momenta variables. [Note that though the Fourier-space integrals to compute are
 in one-to-one correspondence (modulo an overall Fourier transform) with the $\bm x$-space ones which enter the other approach, the computations are somewhat different,
 and the number of integrations to perform over intermediate points in the  $\bm x$-space approach is generally not equal to the
 number of topological loops in the diagram.]

Let us discuss the equivalence between the two approaches in further detail, and also emphasize why it is useful to 
 define the Green's function ${\cal G}(x,y)$ associated with the kinetic operator $ \int [\frac12 \varphi {\cal K } \varphi] $
 as being {\it minus} the inverse of the kinetic term, say
\be \label{green}
{\cal K }\,  {\cal G}(x,y)= -\delta(x-y).
\ee
This was the convention of \cite{Damour:1995kt}, and it leads, when coupling the field to a source $s(x)$, (i.e.\ $\int[\frac12 \varphi {\cal K } \varphi + s \varphi]$)
to a leading-order effective action equal to $+\frac12 \int s(x) {\cal G}(x,y) s(y)$.
Actually, the usefulness of the minus sign in the Green's function definition \eqref{green} is hidden in the usual ``quantum" definition
\eqref{phiphi} of the elementary contraction of the field $\varphi$. Indeed, the rhs of Eq.\ \eqref{phiphi} is really $- (\frac{i}{\hbar} {\cal K })^{-1}$,
i.e.\ minus the inverse of the operator appearing in the exponent of the (functional) integral that one is dealing with.
In other words, the imaginary units $i$ that crowd up the EFT computations are irrelevant. The essential point is that we have two
different ways of approximating an integral of the type
\be  \label{saddleint}
Z[s]=e^{ \frac{1}{\epsilon} S_{\rm eff}}=\int D \varphi\, e^{ \frac{1}{\epsilon}( \int [\frac12 \varphi {\cal K } \varphi + \varphi s  + g \varphi^3 + \cdots])},
\ee
where $\epsilon$ is a formal small parameter, and where the functional measure is normalized so that $Z[s=0]=1$. As the perturbative calculation of $  S_{\rm eff}= \epsilon \ln Z[s]$ is a purely algebraic matter,
one can replace the quantum ``small parameter" $\frac{ \hbar}{i}$ by any formally small parameter  $\epsilon$. One can even simplify
the writing by assuming that the small parameter is absorbed in the definition of the quadratic form $\varphi {\cal K } \varphi$,
and of the interaction terms.
Doing so, the classical approximation to the integral \eqref{saddleint} is to use the saddle-point approximation
\be
Z[s] \approx e^{\int [\frac12 \varphi_* {\cal K } \varphi_* + \varphi_* s  + g \varphi_*^3 + \cdots]},
\ee
where $\varphi_*$ is the saddle point, i.e.\ the solution of 
\begin{align}
\label{saddlepoint}
0 &=\delta \int \left[\frac12 \varphi {\cal K } \varphi + \varphi s  + g \varphi^3 + \cdots\right]/ \delta \varphi
\nonumber\\[1ex]
&= {\cal K} \varphi + s + 3 g \varphi^2 + \cdots.
\end{align}
In this approach, one solves the saddle point condition  \eqref{saddlepoint} by a perturbative series away from the unperturbed solution $\varphi=0$,
namely  [with $ {\cal K }= - {\cal G}^{-1}$ according to the definition \eqref{green}]
\be
\varphi_*= {\cal G}  s + {\cal G} \left(  3 g \, ({\cal G}  s)^2 \right) + \cdots,
\ee
where the needed integrations over intermediate spacetime points are left implicit.
This leads to an expansion of the effective action in powers of the source $s$:
\be \label{seffclassical}
S_{\rm eff}[s] \approx \ln Z[s]^{\rm saddle} \approx \frac12 s  {\cal G}  s + g  \, ({\cal G}  s)^3 + \cdots.
\ee
In the other, Feynman-like approach one approximates (at the exponential accuracy) the integral \eqref{saddleint} by expanding the
integrand away from the Gaussian term 
\be
Z[s]=\int D \varphi\, e^{ \frac{1}{\epsilon}( \int \frac12 \varphi {\cal K } \varphi )} \sum_n \frac{1}{n!} \left( \int[\varphi s  + g \varphi^3 + \cdots]\right)^n,
\ee
using the elementary contraction
\be
\langle \varphi(x) \varphi(y) \rangle = -{\cal K }^{-1}_{x,y}= {\cal G}(x,y).
\ee
From the above reasoning, it is guaranteed that this will give the same result, \eqref{seffclassical}, for the logarithm of $Z[s]$.
But this reasoning shows that all the $i$'s are a useless complication (which can easily lead to sign errors when there are
many of them), as we are computing a real effective action (when using the time-symmetric Green function appropriate
to describing the conservative dynamics).

\section{Explicit expressions of the relevant four-loop, 4PN effective-action contributions}

We focus, in this paper, on the few effective-action contributions that Ref.\ \cite{Foffa:2016rgu} emphasized as being potentially problematic.
As explained in \cite{Foffa:2016rgu} these terms are purely ``static" and follow from the simplified particle + field action
\be
S=S_{\rm pm } + S_{\rm field},
\ee
where the (static) point-mass action is
\begin{align}
S_{\rm pm } &=- \sum_a m_a \int \md t\, e^{\lambda \phi}
\nonumber\\
&= - \sum_a m_a \int \md t\, (1 + \lambda \phi + \cdots),
\end{align}
and where the field action \cite{Kol:2010si,Foffa:2016rgu} is
\begin{align}
\label{sfield}
S_{\rm field} &= \int \md t \md^{d}x \sqrt{\gamma} \bigg[ \frac14 \gamma^{ij} \gamma^{kl} \gamma^{mn} (\D_i \sigma_{kl}  \D_j \sigma_{mn} 
\nonumber \\
&\quad - 2 \D_i \sigma_{km}  \D_j \sigma_{ln}  ) - c_d \gamma^{ij} \D_i \phi \D_j \phi
\nonumber \\
&\quad + \lambda \left(\sigma_{ij} -\frac12 \sigma \delta_{ij}\right) (\sigma_{ik,l} \sigma_{jl,k} - \sigma_{ik,k} \sigma_{jl,l}
\nonumber \\
&\quad + \sigma_{,i} \sigma_{jl,l} - \sigma_{ik,j} \sigma_{,k}) \bigg].
\end{align}
Here, we followed the notation of \cite{Foffa:2016rgu}, apart from the fact that we use $\lambda= 1/\Lambda= \sqrt{32 \pi G \ell_0^{d-3}}$.
The gravitational field degrees of freedom are described by $\phi$ and $\sigma_{ij}$, with $\gamma_{ij}= \delta_{ij} + \lambda \sigma_{ij}$.
In addition, $c_d \equiv 2 \frac{d-1}{d-2}$, $\gamma= \det \gamma_{ij}$,  $\sigma= \sigma_{ii}$, and $f_{,i} \equiv \D_i f$.
Note that, in this approximation, only $\phi$ is directly coupled to the particles. The tensor field $\sigma_{ij}$ is only excited through the cubic vertex following from the
kinetic term of $\phi$:
\begin{align}
\label{kineticphi}
&- c_d  \sqrt{\gamma} \gamma^{ij} \D_i \phi \D_j \phi
\nonumber\\
&\qquad = - c_d \left(\delta_{ij} -  \lambda \sigma_{ij}+ \frac12 \lambda \sigma \delta_{ij} + O(\sigma^2)\right) \D_i \phi \D_j \phi.
\end{align}
For the four-loop terms we are interested in, only the {\it linear} coupling of $\phi$ to the particles, 
\be
\int \md^d x \phi({\bm x}) s({\bm x})  \equiv  -  \sum_a m_a \lambda   \phi({\bm x}_a),
 \ee
matters.  Here the Lagrangian density of the source is
\be \label{source}
s({\bm x}) = - \lambda m_1 \delta_1 - \lambda m_2 \delta_2,
\ee
where $\delta_a \equiv \delta({\bm x} - {\bm x}_a)$.

We can then describe the {\it algebraic} structure of the relevant particle + field Lagrangian as
\be
{\cal L}= {\cal L}_0 + \epsilon {\cal L}_1 ,
\ee
where
\be \label{L0}
{\cal L}_0=-\frac12 \frac{\phi^2}{\cG_{\phi}}  -\frac12 \frac{\sigma^2}{\cG_{\sigma}} + \phi s + a \sigma \phi^2
\ee
includes the kinetic terms, the linear coupling to matter, and the cubic vertex between $\sigma_{ij}$ and $\phi$ coming from
the $\phi$ kinetic term \eqref{kineticphi}, namely
\be \label{cubic}
a \sigma \phi^2 = {\cal L}_{\rm cubic} =  \lambda c_d \left(\sigma_{ij}- \frac12 \sigma \delta_{ij}\right)\D_i \phi \D_j \phi.
\ee
[Note that, following Eq.\ \eqref{green}, we have expressed the kinetic operators of
$\phi$ and $\sigma$ in terms of the corresponding Green's functions $\cG_{\phi}, \cG_{\sigma}$.]
The remaining, higher-order terms in the relevant 4PN action have the algebraic structure
\be \label{L1}
\epsilon {\cal L}_1= b \, \sigma^2 \phi^2 + c \, \sigma^3.
\ee
They respectively correspond to the $O(\sigma^2) \D_i \phi \D_j \phi $ terms in the $\phi$ kinetic term \eqref{kineticphi}, and to the 
sum of the last line in the field action \eqref{sfield}, and of the terms coming from the kinetic terms of $\sigma_{ij}$ when considering
the terms of order $\lambda \sigma_{..}$ in the expansion of 
\be
 \sqrt{\gamma} \gamma^{ij} \gamma^{kl} \gamma^{mn} = \delta_{ij} \delta_{kl} \delta_{mn} + O( \lambda \sigma_{})_{ijklmn},
\ee
using $\sqrt{\gamma} = 1 + \frac12 \lambda  \sigma + O(\lambda^2)$, $ \gamma^{ij} = \delta_{ij} - \lambda \sigma_{ij} + O(\lambda^2)$. Hence,
\begin{align}
\label{csigma3}
c \, \sigma^3 =& \frac14 O( \lambda \sigma_{})_{ijklmn} (\D_i \sigma_{kl}  \D_j \sigma_{mn} - 2 \D_i \sigma_{km}  \D_j \sigma_{ln}  )
\nonumber \\
&\quad + \lambda \left(\sigma_{ij} -\frac12 \sigma \delta_{ij}\right) (\sigma_{ik,l} \sigma_{jl,k} - \sigma_{ik,k} \sigma_{jl,l}
\nonumber \\
&\quad + \sigma_{,i} \sigma_{jl,l} - \sigma_{ik,j} \sigma_{,k}).
\end{align}
As for the terms $ b \, \sigma^2 \phi^2$, they are explicitly given by
\be \label{bsigma2phi2}
b \, \sigma^2 \phi^2 = - c_d \left[ \sqrt{\gamma} \gamma^{ij} \right]_{\sigma^2}  \D_i \phi \D_j \phi,
\ee
with
\begin{align}
\left[ \sqrt{\gamma} \gamma^{ij} \right]_{\sigma^2}
&= \lambda^2 \bigg( \frac18 \sigma^2 \delta_{ij} -\frac14 \sigma_{kl} \sigma_{kl} \delta_{ij}
\nonumber\\
&\quad -\frac12 \sigma \sigma_{ij} + \sigma_{ik} \sigma_{jk}\bigg).
\end{align}
The saddle-point conditions (or field equations of motion) for $\phi$ and $\sigma$ have the structure
\begin{align}
\label{eomphi}
- \frac{\phi}{\cG_{\phi}} + s + 2 a \sigma \phi + \epsilon \frac{\delta \cL_1 }{\delta \phi}&=0,
\\
\label{eomsigma}
- \frac{\sigma}{\cG_{\sigma}} + a \phi^2 + \epsilon \frac{\delta \cL_1 }{\delta \sigma}&=0.
\end{align}
As the solution of these field equations of motion is only needed for being replaced in the Lagrangian $\cL (\phi, \sigma,s)$,
it is well-known that it is enough to solve the equations of motion coming from $\cL_0$, i.e.\ to take $\epsilon=0$ in the above
field equations. Indeed, as $\delta \cL/\delta\,{\rm field}=0$, the corrections to the field solution coming from $\epsilon \cL_1$ contribute only
at order $\epsilon^2$ to the Fokker action. [It is essentially this basic fact that, upon the suggestion one of us (TD), was used to simplify the
recent 4PN harmonic-coordinates computation of the Fokker action \cite{Bernard:2016wrg}.] To lowest-order in a non-linearity
expansion in the source [i.e.\ in an expansion in powers of the two masses $m_1, m_2$, see \eqref{source}], we immediately see
that the solutions of the above field equations are
\begin{subequations}
\label{LOsolutions}
\begin{align}
\phi_* &=  \cG_{\phi} s + O(s^2) ,
\\[1ex]
\sigma_* &= \cG_{\sigma} ( a \phi_*^2) + \cdots=  \cG_{\sigma} ( a ( \cG_{\phi} s )^2) +O(s^4).
\end{align}
\end{subequations}
From the above reasoning, we deduce the first result that the contribution of the action correction $\epsilon {\cal L}_1$ to
the effective (Fokker) action is simply obtained by replacing in $\epsilon {\cal L}_1$ the fields $\phi$ and $\sigma$ by their
{\it lowest-order solution} (because this is enough to get $\epsilon {\cal L}_1$ to order $s^6$), namely
\be
\epsilon {\cal L}_1^{\rm eff} = \Big[ b \, \sigma^2 \phi^2 + c \, \sigma^3 \Big]^{\phi\to \cG_{\phi} s }_{\sigma \to \cG_{\sigma} ( a ( \cG_{\phi} s )^2)}
\ee
This result takes care of two of the contentious action contributions highlighted by \cite{Foffa:2016rgu}, namely $L_{33}$, linked
to $b \, \sigma^2 \phi^2$, and $L_{50}$, linked to $c \, \sigma^3$, and allows one to compute them straightforwardly (including
all numerical factors). The remaining contentious action contribution, $L_{49}$, is easily seen (from its diagram in Fig. 1 of \cite{Foffa:2016rgu}; see also below)
to arise from the exchange of {\it two} cubic vertices, \eqref{cubic}. Therefore, in a Fokker-type
calculation, this term arise from solving the field equations of motion Eqs.\ \eqref{eomphi}, \eqref{eomsigma} to {\it fourth order}
in the  $\phi$-$\sigma$ coupling \eqref{cubic}, that we had left in the zeroth order action $\cL_0$, \eqref{L0}.

\begin{figure}
\includegraphics[scale=0.5]{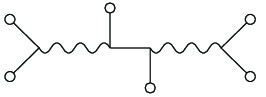}
\caption{\label{fig:a4s6}
The diagrammatic representation of the $O(a^4 s^6)$ contribution to the effective action.}
\end{figure}

It is fairly easy to solve Eqs.\ \eqref{eomphi}, \eqref{eomsigma} (without the $\epsilon \cL_1$ terms) to order $O(a^4)$.
First, let us note that we are talking here about a purely algebraic calculation that could be done by iterating polynomial
expressions. The aim of our calculation is to get the correct numerical coefficient in front of the $O(a^4)$ Fokker action
contribution. This can be formally done by solving Eqs.\ \eqref{eomphi}, \eqref{eomsigma} {\it as if} $\phi$ and $\sigma$
were ordinary numbers. As Eq.\ \eqref{eomphi} (without the $\epsilon \cL_1$ term) is linear in $\phi$ we can solve $\phi$
in terms of $\sigma$ and replace the answer in the second equation. Denoting
\be
x= a \, \cG_{\phi} \sigma,
\quad
x_0= a^2 \, \cG_{\sigma}  \cG_{\phi}^3  s^2,
\ee
the solution of Eq.\ \eqref{eomphi} reads $\phi(\sigma)= (1 - 2 x)^{-1} \cG_{\phi} s$, and its insertion in 
Eq.\ \eqref{eomsigma} (without the $\epsilon \cL_1$ term) reads
\be
x (1-2x)^2 = x_0.
\ee
This is easily solved by iteration in powers of the source:
\be
x= x_0( 1 + 4 x_0 + O(x_0^2)).
\ee
Inserting this solution in $\cL_0$ then easily leads to an expansion in {\it even} powers of $s$:
\be
\cL_0^{\rm eff}[s]= \frac12  \cG_{\phi} s^2 +\frac12 a^2 \cG_{\sigma}  \cG_{\phi}^4  s^4+ 2 \, a^4 \cG_{\sigma}^2  \cG_{\phi}^7  s^6 +O(s^8).
\ee
Here, we are interested in the third term of order $s^6$, i.e.\ involving six masses. The aim of the above algebraic calculation was to
safely derive the numerical factor in front of this contribution (which is linked to $L_{49}$). It is easy to understand which diagram
this term is connected with by rewriting it as (denoting the linear-in-source solution as $\phi_*^{(1)} \equiv \cG_{\phi} s $)
\be
\frac12  [2 a [ a \cG_{\sigma} (\phi_*^{(1)})^2 ] \phi_*^{(1)}]  [\cG_{\phi}] [2 a [ a \cG_{\sigma} (\phi_*^{(1)})^2] \phi_*^{(1)}],
\ee
where the nested brackets on each side (starting with $[2a [\cdots] \cdots]$) denote the third-order (in $s$) solution of the
$\phi$ equation, i.e.\ the second term, $\phi_*^{(3)}$, in
\bea
\phi_*&=& \phi_*^{(1)}+ \phi_*^{(3)}= \cG_{\phi} s + \cG_{\phi} [2 a [ a \cG_{\sigma} (\phi_*^{(1)})^2 ] \phi_*^{(1)}]
\nonumber \\
&=&  \cG_{\phi} s +  \cG_{\phi} [ 2 a \, \sigma_*^{(2)} \phi_*^{(1)}],
\eea
where 
\be
\sigma_*^{(2)} =  \cG_{\sigma} \left( a \, (\phi_*^{(1)})^2  \right)
\ee
 denotes the lowest-order (quadratic in $s$) solution for $\sigma$,
obtained by inserting  $ \phi_*^{(1)}$ in the effective source ($a\, \phi^2$) of $\sigma$ [see Eq.\ \eqref{LOsolutions}].
The diagrammatic representation of this $O(a^4 s^6)$ contribution to the effective action is displayed in Fig.\ \ref{fig:a4s6}.

A useful way of reexpressing the $O(a^4 s^6)$ contribution to $\cL_0^{\rm eff}[s]$ is to write it as
\be \label{a4s6}
\left[\cL_0^{\rm eff}[s] \right]_{a^4 s^6}= \frac12 \left[\frac{\delta \cL_{\rm cubic}}{\delta \phi}\right]_{\rm LO} \phi_*^{(3)},
\ee
where $\cL_{\rm cubic}= a \sigma \phi^2$ is the cubic $\sigma$-$\phi$ coupling, Eq.\ \eqref{cubic}, and where all the
fields in $\delta \cL_{\rm cubic}/{\delta \phi }= 2 a \, \sigma \phi$ on the rhs can be replaced by their lowest-order solutions.

\section{Explicit computation of the contentious four-loop, 4PN effective-action contributions}

We have given in the preceding section all the material needed to write down, in ${\bm x}$-space, all the integrals $L_{26}$ to
$L_{50}$ in Fig.\ 1 of  \cite{Foffa:2016rgu}, i.e.\ all the $O(s^6)$ diagrams where the $\phi$ field couples
only linearly to the particles. [The other $O(s^6)$ diagrams $L_{1}$ to $L_{25 }$ in Fig.\ 1 of  \cite{Foffa:2016rgu} all involve some $\phi^n \cdot s$
coupling with $n \geq 2$.]

Among the integrals $L_{26}$ to $L_{50}$, we are only interested in reevaluating the three integrals 
${ L}_{33}$, ${ L}_{49}$ and ${ L}_{50}$, which contain the transcendental coefficient $\zeta(2)=\pi^2/6$, and whose evaluations
in Ref.\ \cite{Foffa:2016rgu} gave the problematic values \eqref{fmss}. The method of
computation used in Ref.\ \cite{Foffa:2016rgu} was the Feynman-like one sketched above: in ${\bm p}$-space, with purely imaginary
propagators $ i {\cal K}^{-1}$, and with the use of integration by parts identities to reduce the multi-loop  ${\bm p}$-space
integrals to a subset of master integrals [one of them, ${\cal M}_{3,6}$ could only be evaluated numerically, though with
such a high accuracy that they could recognize the presence of $\zeta(2)=\pi^2/6$ in it].

In the following three subsections we shall reevaluate the four-loop integrals ${ L}_{33}$, ${ L}_{49}$ and ${ L}_{50}$, 
in ${\bm x}$-space, using  ${\bm x}$-space integration by parts, and using as master integrals only the ones that have been
used in our previous PN (and ADM) work, namely the original Riesz integration formula \cite{Riesz1949} [which was crucially used in the
first complete computation of the (harmonic-coordinates) 2PN action (containing up to two-loop diagrams) \cite{Damour1982,Damour1983}],
together with the ``generalized Riesz formula" (first derived in \cite{Jaranowski:1997ky} for the computation
of the 3PN Hamiltonian, and which was also sufficient for the computation of the local ADM 4PN 
Hamiltonian computation \cite{Jaranowski:2015lha}). To streamline the presentation of our computations, we will
relegate most of the needed, general integration formulas to Appendix~\ref{appendixA}.

\subsection{$L_{33}$}

In ${\bm x}$-space, $L_{33}$ arises (together with its cousins $L_{26}, L_{27}, L_{28}, L_{29}, L_{30}, L_{31}, L_{32}$,
and $L_{34}$ in Fig.\ 1 of  Ref.\ \cite{Foffa:2016rgu}) from an integral of the form
\be
\label{L33a}
L_{\sigma \sigma \phi \phi} = \int \md^d x \, \sigma  \sigma \D \phi \, \D \phi,
\ee
in which $\phi$ and $\sigma$ must be replaced by their lowest-order solutions, denoted $\phi_*^{(1)}$ and $ \sigma_*^{(2)}$
above, so that $L_{\sigma \sigma \phi \phi}$ is of sixth order in the masses. The explicit expression of the integrand 
$\sigma \sigma \D \phi \, \D \phi $ is obtained from inserting Eq.\ (3.14) in Eq.\ (3.13), and reads
\begin{align}
\label{L33b}
\sigma  \sigma \D \phi \, \D \phi &= -\lambda^2 c_d
\bigg( \frac18 \sigma^2 \delta_{ij} -\frac14 \sigma_{kl} \sigma_{kl} \delta_{ij}
\nonumber\\
&\quad  -\frac12 \sigma \sigma_{ij} + \sigma_{ik} \sigma_{jk}\bigg)
 \D_i\phi \, \D_j\phi.
\end{align}
When decomposing $\phi_*^{(1)}$ and $ \sigma_*^{(2)}$ according to their mass content, i.e.
\be
\phi_*^{(1)}= m_1 \phi_1+ m_2  \phi_2
\ee
and
\be
\sigma_*^{(2)} = m_1^2 \sigma_{11} + m_1 m_2 \sigma_{12} + m_2^2 \sigma_{22},
\ee
one recovers all the diagrams $L_{26}$ to  $L_{27}$  (modulo some vanishing self-gravity ones, and the
$ 1\leftrightarrow 2$ images of the previous ones).

\begin{figure}
\includegraphics[scale=1.1]{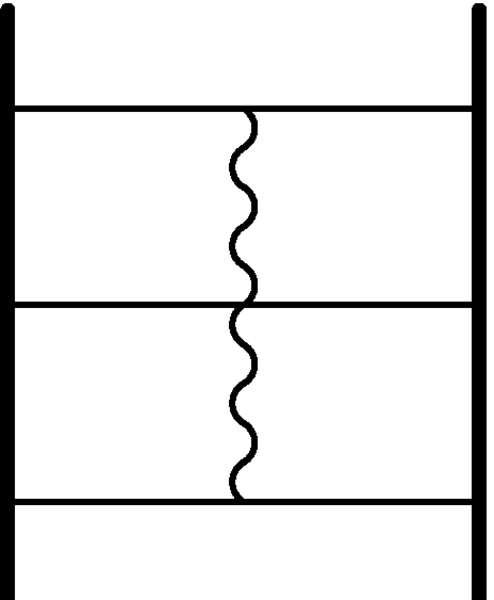}
\caption{\label{L33spacetime}
The spacetime diagram of $L_{33}$.}
\end{figure}

\begin{figure}
\includegraphics[scale=1]{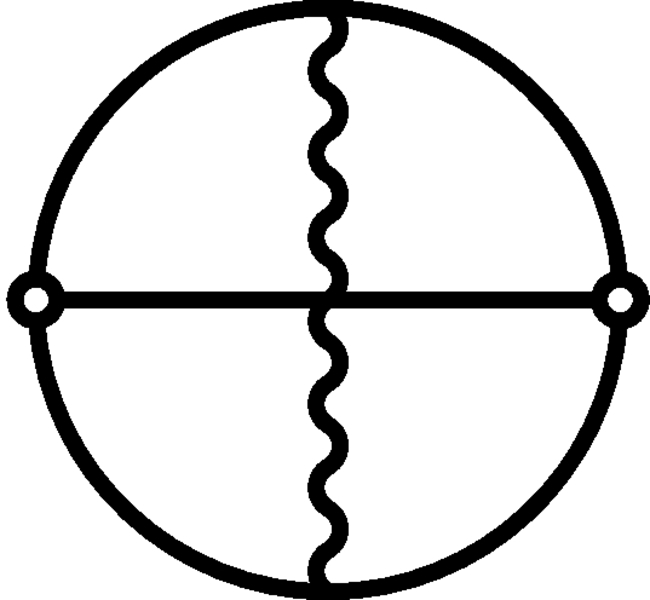}
\caption{\label{L33space}
The spatial projection of the diagram of $L_{33}$.}
\end{figure}

But we are only interested in $L_{33}$, given by the spacetime diagram Fig.\ \ref{L33spacetime}. 
(The thin lines represent the $\phi$ propagators, while the wavy lines represent the $\sigma$ propagators.)
The $d$-dimensional projection of the diagram of $L_{33}$ is the two-point, four-loop diagram Fig.\ \ref{L33space}.
(Here, the empty circles represent the two point-mass sources, i.e.\ the spatial projections of the thick, external worldlines
in the corresponding spacetime diagram.)
We see on its representation that this diagram (modulo the convention $\cL_{33}^{\rm FMSS}= \frac12 L_{33}^{\rm here}$)
is obtained from the general integral $L_{\sigma \sigma \phi \phi} $ by replacing each $\sigma$ by $m_1 m_2 \sigma_{12}$,
and one $\phi$ by $m_1 \phi_1$ and the other by $m_2  \phi_2$, so that
\be
\label{L33}
L_{33}= 2 \, m_1^3 m_2^3 \, L_{\sigma_{12} \sigma_{12} \phi_1 \phi_2 },
\ee
the factor 2 taking into account the two possibilities $\phi_1 \phi_2$ vs $\phi_2 \phi_1$.

We explained in Sec.\ III above the definitions of $\phi_*^{(1)}$ and $\sigma_*^{(2)} $ in terms of sources and propagators.
In practical terms, the consideration of the Euler-Lagrange equations defined by the action \eqref{sfield} yields
\be
\Delta \phi= \frac{\lambda}{2 \, c_d} \sum_a m_a \delta_a + \cdots,
\ee
so that (using standard $d$-dimensional formulas recalled in Appendix \ref{appendixA})
\be
\phi_*^{(1)}= - \frac{\tilde k}{4 \pi}  \frac{\lambda}{2 \, c_d} \sum_a m_a r_a^{2-d},
\ee
where $r_a \equiv |{\bm x} - {\bm x}_a |$.

Writing the field equation for $\sigma_{ij}$ following from the action \eqref{sfield} yields (after a simple manipulation)
\be
\Delta \sigma_{ij} = - \lambda c_d \D_i \phi \D_j \phi + \cdots,
\ee
so that
\be
\sigma_{* ij}^{(2)} =  - \lambda c_d \Delta^{-1} [\D_i \phi_*^{(1)} \D_j \phi_*^{(1)} ].
\ee
In particular, we see that the mixed contribution $m_1 m_2 \sigma_{12}$ to $\sigma_{* ij}^{(2)}$ can be expressed
(in ${\bm x}$-space) in terms of partial derivatives (with respect to  ${\bm x}_1$ and ${\bm x}_2$) of the $d$-dimensional 
potential $g_d$ defined by
\be
g_d({\bm x}, {\bm x}_1, {\bm x}_2) \equiv  \Delta^{-1}( r_1^{2-d} r_2^{2-d}).
\ee
An explicit expression for $g_d({\bm x}, {\bm x}_1, {\bm x}_2)$ was derived in Appendix C of \cite{Blanchet:2003gy}.
Let us only recall now that, when $\e = d-3 \to 0$, one has the formal result
\be
g_d= - \frac1{2 \, \e (1-\e)} + \ln \left(\frac{r_1+r_2+r_{12}}{2} \right) + O(\e),
\ee
so that one recovers the well-known fact (originally due to Fock \cite{Fock1955}) that, in {\it three} dimensions,
\be
{\bar g}_3 \equiv  \ln \left(\frac{r_1+r_2+r_{12}}{2} \right) 
\ee
is a solution of $\Delta  {\bar g}_3= r_1^{-1} r_2^{-1} $.
(It will be convenient in the following to include the
factor $\frac12$ in the argument of the logarithm.)

In three dimensions, the explicit expression of  $ \sigma_{12}$ is
\be
\label{sigma12}
\sigma_{12}= - \frac1{(4\pi)^2} \frac{\lambda^3}{16} \left(  \D_i^1 \D_j^2  {\bar g}_3 +  \D_j^1 \D_i^2  {\bar g}_3 \right).
\ee
where $ \D_i^a \equiv \D/\D x^i_a$ ($a=1,2$), while the functions $\phi_a$ ($a=1,2$) read
\be
\label{phia}
\phi_a = - \frac{\lambda}{32\pi}\frac{1}{r_a}.
\ee
It is then easily seen that, in $d=3$, the integral $L_{33}$ is convergent {\it both} in the ultraviolet (UV),
i.e.\ near the point masses, and in the infrared (IR), i.e.\ at spatial infinity. 

In three dimensions the integrand of $L_{33}$
[see Eq.\ \eqref{L33} together with Eqs.\ \eqref{L33a}--\eqref{L33b} and \eqref{sigma12}--\eqref{phia}]
can be explicitly  written as
\begin{widetext}
\begin{align}
\frac{\lambda^{10}m_1^2m_2^3}{(16\pi)^6}
\frac{\big[(r_1 - r_2)^2-r_{12}^2\big] \big[(r_1 - r_{12}) (r_1 + r_{12})^3 - 
   2 r_{12}^3 r_2 - 2 r_1^2 r_2^2 + 2 r_{12} r_2^3 + r_2^4\big]}{2 r_1^4 r_{12}^4 r_2^4 (r_1 + r_2 + r_{12})^2},
\end{align}
\end{widetext}
so that, for evaluating the integral $L_{33}$, it is enough to use the generalized Riesz formula $\hat{I}[a,b,c]$ with $c=-2$
(see Appendix \ref{appendixA}).
Our final result is
\be
L_{33} =  (32 - 2 \pi^2)  \frac{G^5 m_1^3 m_2^3}{c^8 r_{12}^5}.
\ee

\subsection{$L_{49}$}

\begin{figure}
\includegraphics[scale=1.2]{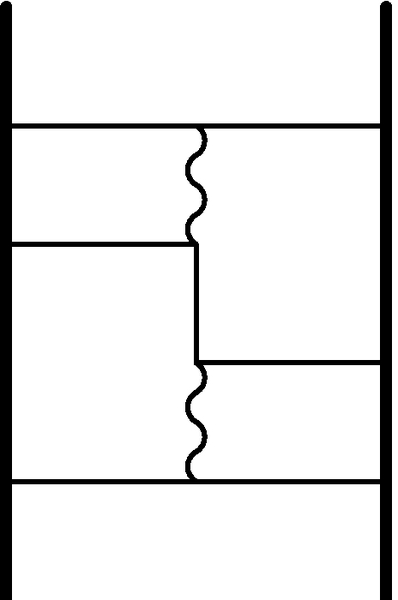}
\caption{\label{L49spacetime}
The spacetime diagram of $L_{49}$.}
\end{figure}

\begin{figure}
\includegraphics[scale=1]{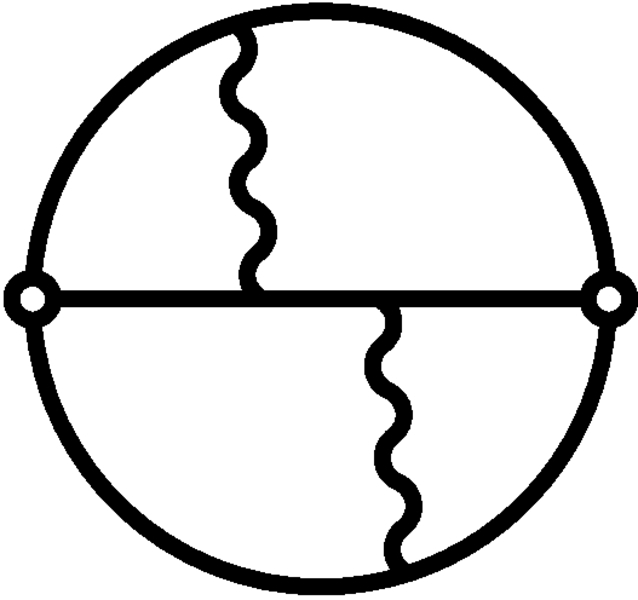}
\caption{\label{L49space}
The spatial projection of the diagram of $L_{49}$.}
\end{figure}

In ${\bm x}$-space, $L_{49}$ arises (together with its cousins $L_{35}, L_{36}, L_{37}, L_{38}, L_{39}, L_{41}, L_{42}, L_{44}, L_{47}$,
and $L_{48}$ in Fig.\ 1 of  Ref.\ \cite{Foffa:2016rgu}) from the $O(a^4 s^6)$ contribution  to $\cL_0^{\rm eff}[s]$
given by Eq.\ \eqref{a4s6}. The corresponding spacetime diagram is displayed on Fig.\ \ref{L49spacetime},
while its (two-point, four-loop) spatial projection is shown on Fig.\ \ref{L49space}.

Using the fact that the explicit expression of the $\sigma$-$\phi^2$ cubic vertex is given by Eq.\ \eqref{cubic},
it is easily seen (after an integration by parts) that the latter contribution can be written as
\be
\frac12 \left[\frac{\delta \cL_{\rm cubic}}{\delta \phi}\right]_{\rm LO} \phi_*^{(3)}
= - \lambda^2 c_d \int \md^d x \, \omega \Delta^{-1} \omega,
\ee
where
\be
\omega \equiv \D_i \left[ \left( \sigma_{ij}-\frac12 \sigma \delta_{ij} \right)   \D_j \phi  \right],
\ee
in which $\phi$ and $\sigma$ must be replaced by their lowest-order solutions, denoted $\phi_*^{(1)}$ and $ \sigma_*^{(2)}$
above. From the form of the diagram of $L_{49}$ one sees that one must keep in $\omega$ only the two pieces generated
by $m_1 m_2 \sigma_{12}$ and bilinear in the two pieces of $\phi_*^{(1)}= m_1 \phi_1 + m_2 \phi_2$. Defining
\be \label{omega12a}
\omega^{12}_a \equiv \D_i \left[ \left( \sigma_{ij}^{12}-\frac12 \sigma^{12} \delta_{ij} \right)   \D_j \phi_a  \right],
\ee
where, for clariy, we put the mass labels $12$ of $\sigma$ as superscripts, we end up with
\be
L_{49}=  - 2 \,  \lambda^2 c_d \, m_1^3 m_2^3  \int \md^d x \, \omega^{12}_1 \Delta^{-1} \omega^{12}_2,
\ee
where the extra factor $2$ takes into account the two orderings $\omega^{12}_1 \omega^{12}_2$ vs $\omega^{12}_2 \omega^{12}_1$.

The integral $L_{49}$ is IR convergent, but it has a mildly singular UV behavior because of the presence
of two derivatives of $\phi_a$ in $\omega^{12}_a$ (when expanding its definition \eqref{omega12a}).
One must treat these derivatives in a distribution-theory way.
After evaluating all differentiations present in $\omega_1^{12}$ one gets (in $d=3$)
\be
\label{omega1a}
\omega_1^{12} = \omega_{1\,\text{fun}}^{12} + \omega_{1\,\text{DD}}^{12},
\ee
where
\begin{subequations}
\label{omega1b}
\begin{align}
\omega_{1\,\text{fun}}^{12} &= \frac{\lambda^4 }{(16\pi)^3}
\bigg[ \frac{3}{4}\left(\frac{r_{12}}{r_2} - \frac{2 r_2}{r_{12}} +\frac{ r_2^3}{r_{12}^3}\right)\frac{1}{r_1^5}
\nonumber\\[1ex]
&\quad - \frac{1}{2} \left(\frac{1}{r_2} - \frac{2}{r_{12}} + \frac{r_2^2}{r_{12}^3}\right) \frac{1}{r_1^4}
- \frac{r_2}{r_{12}^3 r_1^3}
\nonumber\\[1ex]
&\quad + \frac{1}{2 r_{12}^3 r_1^2} + \frac{1}{4 r_{12}^3 r_1 r_2} \bigg],
\\[1ex]
\omega_{1\,\text{DD}}^{12} &= -\frac{\pi}{3r_{12}^2}\frac{\lambda^4 }{(16\pi)^3}\delta_1.
\end{align}
\end{subequations}
It is not difficult to find the function $\chi_1$ such that $\Delta\chi_1 = \omega_{1\,\text{fun}}^{12}$ in the sense of functions.
It reads
\begin{align}
\chi_1 &= \frac{\lambda^4 }{(16\pi)^3}\bigg[
\frac{1}{4}\left(\frac{r_2}{r_{12}} - \frac{r_2^3}{r_{12}^3}\right) \frac{1}{r_1^3}
\nonumber\\[1ex]
&\quad + \frac{1}{4}\left(\frac{r_2^2}{r_{12}^3} - \frac{r_2}{r_{12}^2}\right) \frac{1}{r_1^2} + \frac{r_2}{4 r_{12}^3}\frac{1}{r_1}
\bigg].
 \end{align}
 Computation of $\Delta\chi_1$ in the sense of distributions gives extra distibutional terms
 \be
 \Delta\chi_1 = \omega_{1\,\text{fun}}^{12} + \frac{\lambda^4 }{(16\pi)^3}\left(
 \frac{2\pi}{3r_{12}^2}\delta_1
 - \frac{2\pi}{r_{12}}{\bm n}_{12}\cdot\nabla\delta_1 \right).
 \ee
 Hence, in the sense of distributions,
 \be
 \Delta\left[\chi_1 + \frac{\lambda^4 }{(16\pi)^3}\left(
 \frac{1}{6r_{12}^2}\frac{1}{r_1}
 - \frac{1}{2r_{12}}{\bm n}_{12}\cdot\nabla\frac{1}{r_1} \right)\right] = \omega_{1\,\text{fun}}^{12}.
\ee
Taking this result into account as an inverse Laplacian of $\omega_1$ we take
\begin{widetext}
\begin{align}
\label{invomega1}
\Delta^{-1}\omega_1^{12} &= \chi_1 + \frac{\lambda^4 }{(16\pi)^3}\left(
\frac{1}{6r_{12}^2}\frac{1}{r_1}
- \frac{1}{2r_{12}}{\bm n}_{12}\cdot\nabla\frac{1}{r_1} \right)
+  \Delta^{-1} \omega_{1\,\text{DD}}^{12}
\nonumber\\[1ex]
&= -\frac{\lambda^4 }{4(16\pi)^3}\left[
\left(1 - \frac{r_2}{r_{12}} - \frac{r_2^2}{r_{12}^2} + \frac{r_2^3}{r_{12}^3}\right) \frac{1}{r_1^3}
+ \frac{1}{r_{12}} \left(\frac{r_2}{r_{12}} - \frac{r_2^2}{r_{12}^2}\right) \frac{1}{r_1^2}
- \frac{r_2}{r_{12}^3} \frac{1}{r_1}
\right].
\end{align}
\end{widetext}
Making use of Eqs.\ \eqref{omega1a}--\eqref{omega1b} and \eqref{invomega1},
the integrand of $L_{49}$ can symbolically be written as
\begin{align}
\label{L49integrand}
\lambda^2 m_1^3 m_2^3 \omega^{12}_1 \Delta^{-1} \omega^{12}_2 &\sim \frac{\lambda^{10} m_1^3 m_2^3}{(16\pi)^6} \sum_k d_k r_1^{a_k}r_2^{b_k}r_{12}^{c_k}
\nonumber\\[1ex]
&\quad + \frac{\lambda^{10} m_1^3 m_2^3}{(16\pi)^5} \sum_k d'_k r_1^{a'_k}r_2^{b'_k}r_{12}^{c'_k}\delta_1,
\end{align}
where $a_k$, $b_k$, $c_k$, $a'_k$, $b'_k$, $c'_k$ are integers
and the coefficients $d_k$ and $d'_k$ are rational numbers.
The integral of the first part of \eqref{L49integrand} is evaluated by means of the (ordinary) Riesz formula
while the integral of the second part is computed by using Hadamard partie finie procedure.
Our final result is
\be
L_{49} =  (64 - 6 \pi^2)  \frac{G^5 m_1^3 m_2^3}{c^8 r_{12}^5}.
\ee

\subsection{$L_{50}$}

\begin{figure}
\includegraphics[scale=1.2]{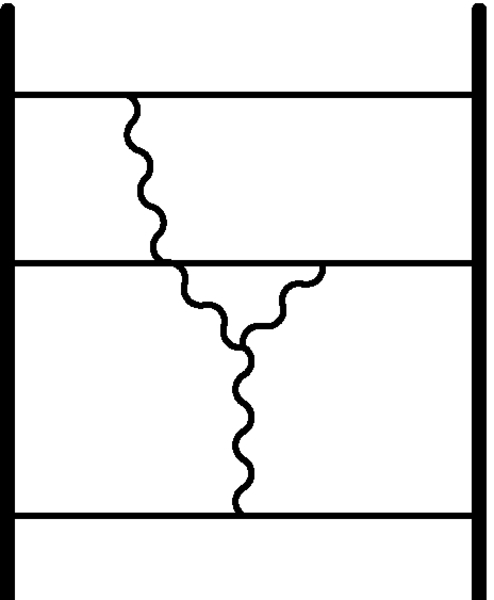}
\caption{\label{L50spacetime}
The spacetime diagram of $L_{50}$.}
\end{figure}

\begin{figure}
\includegraphics[scale=1]{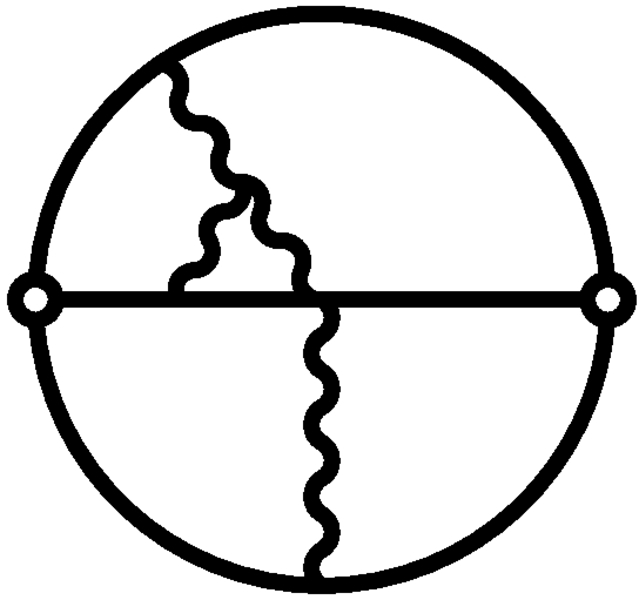}
\caption{\label{L50space}
The spatial projection of the diagram of $L_{50}$.}
\end{figure}

In ${\bm x}$-space, $L_{50}$ arises (together with its cousins $L_{40}$, $L_{43}$, $L_{45}$,
and $L_{46}$ in Fig.\ 1 of  Ref.\ \cite{Foffa:2016rgu}) from the effective action contribution denoted $c \sigma^3$
above, and defined in Eq.\ \eqref{csigma3}.  The spacetime diagram of $L_{50}$ is displayed in Fig.\ \ref{L50spacetime},
while its (non-planar, two-point, four-loop) spatial projection is shown in Fig.\ \ref{L50space}.

Again the term $L_{50}$ we are interested in is, as seen on its
diagram, selected from this cubic expression in $\sigma$ by replacing each occurrence of $\sigma$ by its
mixed piece $m_1 m_2 \sigma_{12}$, i.e., symbolically
\be
L_{50} =   m_1^3 m_2^3 \, \int \md^dx \,  c\, (\sigma_{12})^3
\ee
without any extra symmetry factor. 

The integral $L_{50}$ is {\it both} IR and UV convergent.
In three dimensions its integrand can be symbolically  written as
\be
\frac{\lambda^{10} m_1^3 m_2^3}{(16\pi)^6}
\sum_k d_k \frac{r_1^{a_k}r_2^{b_k}r_{12}^{c_k}}{(r_1+r_2+r_{12})^3},
\ee
where $a_k$, $b_k$, $c_k$ are integers and the coefficients $d_k$ are rational numbers.
For evaluation of the integral $L_{50}$ it is thus enough to use the generalized Riesz formula $\hat{I}[a,b,c]$ with $c=-3$.
Our final result is
\be
L_{50} =  \left(- \frac{248}{3} + 8 \pi^2   \right)  \frac{G^5 m_1^3 m_2^3}{c^8 r_{12}^5}.
\ee

\subsection{Total result, and comparison with Ref.\ \cite{Foffa:2016rgu}}

The crucial result of our new computations is that the transcendental coefficients $\sim \pi^2$
{\it cancell} in the sum of the three contributions $L_{33}$, $L_{49}$, and $L_{50}$:
\be
L_{33} + L_{49} + L_{50}= + \frac{40}{3} \frac{G^5 m_1^3 m_2^3}{c^8 r_{12}^5}.
\ee
This cancellation comes about because, while our results for $L_{33}$, and $L_{49}$ {\it agree} with the
corresponding results of Ref.\ \cite{Foffa:2016rgu} recalled in Eq.\ \eqref{fmss} above, our result for $L_{50}$
differs from the corresponding result of Ref.\ \cite{Foffa:2016rgu} by a factor $-3$:
\be
L_{50}= - 3 \,  L_{50}^{\rm FMSS} = - 6 \, \cL_{50}^{\rm FMSS}.
\ee
It would be interesting to understand the origin of such a missing factor $-3$ in Ref.\ \cite{Foffa:2016rgu}.
It might be caused by the presence of many $i$'s (including the ones linked to the Fourier transform of spatial derivatives
$\D_j \to i p_j$) in the quantum, ${\bm p}$-space calculation of $S_{\rm eff}$, together with
an incorrect account of the pesky symmetry factors that enter any Wick-contraction calculation.

Anyway, we trust our result for $L_{50}$ because its normalization is very straightforwardly obtained
in our ${\bm x}$-space computation. It would be also important to know if the error in $ L_{50}^{\rm FMSS}$
has affected other integrals in Ref.\ \cite{Foffa:2016rgu}. (Because of the cancellation of all the pole parts $\sim \frac1{\e}$
in the genuine $G^5$ contribution the cousins $L_{40}$, $L_{43}$, $L_{45}$, $L_{46}$ of $ L_{50}$ cannot be uniformly affected
by the same factor $-3$.)

Another reason for trusting our results is that they now reconcile the finding announced in \cite{Foffa:2016rgu} that
all the currently known $\pi^2$-dependent coefficients at order $G^5$ in the harmonic-ccordinates version of
$L_{{\text{4PN}}}^{\rm int}\big|^{O(G^5)}_{{\rm linear} \ {\rm in}\ \dot{\bm v}_a}$ come from the double-zero reduction of the
quadratic-in-acceleration terms in the original $O(G^3)$ action, see Eq.\ \eqref{pi2froma2}. The correctness of
the $O(m_1^3 m_2^3)$ sector of the harmonic-coordinates action of \cite{Bernard:2015njp} was strongly 
expected in view of its agreement with the corresponding sector of the ADM action. [In terms of the $\mu$-reduced
Hamiltonian, this corresponds to $O(\nu^2)$ terms that had been unambiguously derived already in Ref.\ \cite{Jaranowski:2013lca}.]

\section{Analytic computation of the master integral $\cM_{3,6}$}

\begin{figure}
\includegraphics[scale=1.2]{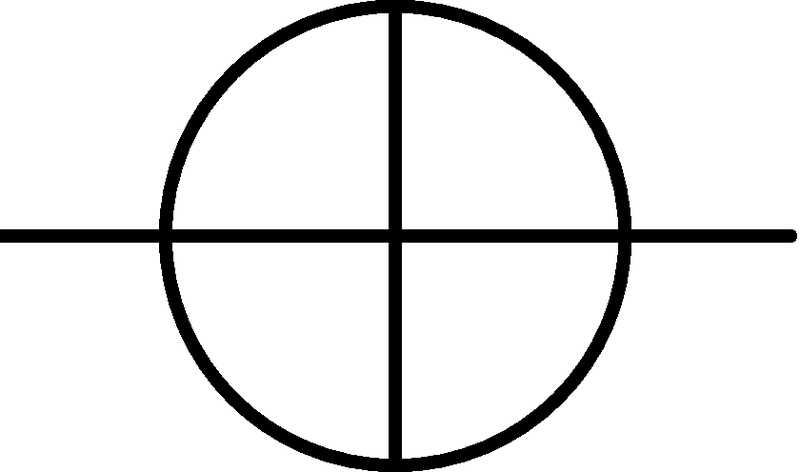}
\caption{\label{fig:m36p}
The master integral $\cM_{3,6}({\bm p})$.}
\end{figure}

\begin{figure}
\includegraphics[scale=1]{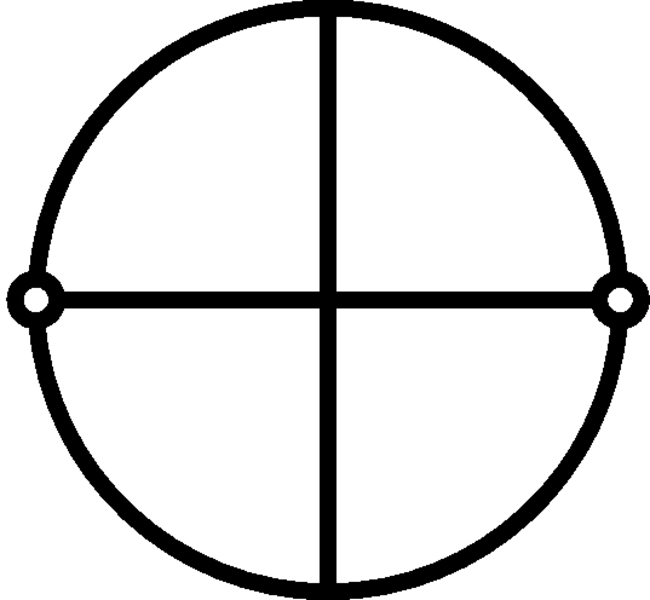}
\caption{\label{fig:u1u2g2}
The ${\bm x}$-space integral $I^{(d)}_{u_1u_2 g^2}$.}
\end{figure}

The master integral denoted $\cM_{3,6}$ in Ref.\ \cite{Foffa:2016rgu}
is the $d$-dimensional ${\bm p}$-space, four-loop integral depicted in Fig.\ \ref{fig:m36p}, and defined by
\be
\label{m36}
\cM_{3,6}({\bm p}) \equiv \int \frac{{\bar \md}k_1\,{\bar \md}k_2\,{\bar \md}k_3\,{\bar \md}k_4}{D_{3,6}},
\ee
where ${\bar \md} k_i \equiv \md^d k_i/(2 \pi)^d$ are normalized Fourier integrals over the loop momenta,
and where the denominator is
\begin{align}
D_{3,6}&= k_1^2 k_2^2 k_3^2 k_4^2 (k_2-k_3)^2 (k_1-k_4)^2
\nonumber\\[1ex]
&\quad \times (k_1+k_2-p)^2 (k_1+k_2-k_3-k_4-p)^2.
\end{align}
Modulo some normalization factors, this is the Fourier transform of the following $d$-dimensional ${\bm x}$-space integral 
\be \label{u1u2g2}
I^{(d)}_{u_1u_2 g^2} \equiv \int \md^d x \, u_1 \, u_2 \, (g_d)^2,
\ee
where
\be
u_1\equiv r_1^{2-d}, \quad u_2\equiv r_2^{2-d}, \quad g_d \equiv \Delta^{-1} (u_1 u_2).
\ee
The diagrammatic representation (in ${\bm x}$-space) of the (scalar, massless) two-point, four-loop integral \eqref{u1u2g2}
is displayed in Fig.\ \ref{fig:u1u2g2}. Note that both the ${\bm p}$-space and ${\bm x}$-space representations of this Feynman integral
have the shape of a four-spoked wheel.

The basic reason why the four-loop integral \eqref{u1u2g2} can be analytically computed near $d=3$ by means of the
generalized Riesz formula is seen in Eq.\ (4.11): near $d=3$, $g_d \equiv \Delta^{-1}(u_1 u_2)$ contains the (Fock) function
$\ln s$, where 
\be
s \equiv r_1 + r_2 + r_{12}.
\ee
Therefore, the integral \eqref{u1u2g2} will contain (near $d=3$) a sum of terms of the type $\int \md^3x\, r_1^{-1} r_2^{-1} (\ln s)^2 $
and $\int \md^3x\,  r_1^{-1} r_2^{-1}\ln s$, which can be obtained by differentiating the generalized Riesz formula with respect to
the exponent of $s$.  However, there are some tricky details when implementing such a computation of \eqref{u1u2g2}, as will be
now explained.

First, one must cope with the IR-divergence of \eqref{m36}, or equivalently \eqref{u1u2g2}. This IR-divergence
is rooted in the IR-divergence of $g_d$ itself, which shows up in the $1/\e$ contribution (where we recall $\e\equiv d-3$) in
Eq.\  (4.11). Let us define
\be
C_0 \equiv \frac1{(2d-6) (4-d)} \equiv \frac1{2 \e (1-\e)}
\ee
and let us consider the new integral
\be
I^{(d)}_{u_1u_2 {\bar g}^2} \equiv \int \md^d x \, u_1 \, u_2 \, ({\bar g}_d)^2,
\ee
where
\be
{\bar g}_d \equiv g_d  +  C_0.
\ee
The latter definition is such that ${\bar g}_d$ has a (point-wize) {\it finite} limit in 3 dimensions, namely
\be
\lim_{\e\to 0} {\bar g}_d ({\bm x}, {\bm x}_1, {\bm x}_2) = \ln \frac{s}{2}.
\ee
We have
\be
\label{g2reduction}
I^{(d)}_{u_1u_2 {\bar g}^2} = I^{(d)}_{u_1u_2 g^2} + 2 \, C_0 \, I^{(d)}_{u_1u_2 g} + C_0^2 \, I^{(d)}_{u_1u_2 },
\ee
where we defined
\be
I^{(d)}_{u_1u_2 g}  \equiv \int \md^d x \, u_1 \, u_2 \, g_d
\ee
and 
\be
I^{(d)}_{u_1u_2 }  \equiv \int \md^d x \, u_1 \, u_2.
\ee
From Eq.\ \eqref{g2reduction}, we see that we can reduce the computation of  $ I^{(d)}_{u_1u_2 g^2}$ to that of the
three integrals: $I^{(d)}_{u_1u_2 {\bar g}^2}$, $ I^{(d)}_{u_1u_2 g}$ and  $I^{(d)}_{u_1u_2}$.
The last integral is trivially given by the standard $d$-dimensional\footnote{Because of the $\e$-singular factors
$C_0\sim \frac1{\e}$ and $C_0^2 \sim \frac1{\e^2}$ one needs to use the values of $ I^{(d)}_{u_1u_2 g}$ and  $I^{(d)}_{u_1u_2}$
in $d$ dimensions.} Riesz integral. After division by $\Omega_d$, where $\Omega_d=2\pi^{d/2}/\Gamma(d/2)$ denotes the surface
of the unit sphere in $d$-dimensional Euclidean space, one finds
\be
\frac1{\Omega_d} I^{(d)}_{u_1u_2} = \frac12 \frac{d-2}{d-4} r_{12}^{4-d}
= -\frac12 \frac{1+\e}{1-\e} r_{12}^{1-\e}.
\ee

\begin{figure}
\includegraphics[scale=1]{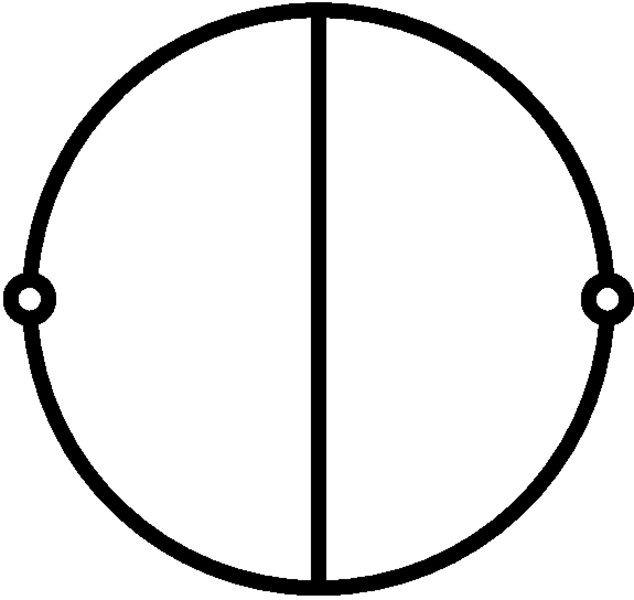}
\caption{\label{fig:u1u2g}
The ${\bm x}$-space integral $I^{(d)}_{u_1 u_2 g}$.}
\end{figure}

The intermediate integral in Eq.\ \eqref{g2reduction}, namely $ I^{(d)}_{u_1u_2 g}$,
is a much simpler integral than $I^{(d)}_{u_1u_2 {\bar g}^2}$
because it is a massless two-loop, two-point (scalar) Green function. It is depicted in Fig.\ \ref{fig:u1u2g}.
Such Green's functions have been computed
in the Feynman-integral literature. More precisely, the Fourier transform of $I^{(d)}_{u_1u_2 g}$ (modulo some
different normalization factors, including an overall sign) has been computed
by Chetyrkin, Kataev, and Tkachov using Gegenbauer-polynomial,  ${\bm x}$-space techniques\footnote{We note in passing that
similar techniques have been used to compute $g_d$ itself in $d$ dimensions \cite{Blanchet:2003gy}, and the generalized
Riesz formula in 3 dimensions \cite{Jaranowski:1997ky}.} \cite{Chetyrkin:1980pr}  (see also \cite{Suzuki:2014hda}).
It is trivial to compute the inverse Fourier transform of the result of Refs.\ \cite{Chetyrkin:1980pr,Suzuki:2014hda}
(given in Appendix \ref{appendixA}), 
so as to compute the exact analytical expression of $ I^{(d)}_{u_1u_2 g}$,
namely
\be
 \frac1{\Omega_d} I^{(d)}_{u_1u_2 g}= N^{(d)}_{u_1u_2  g} \ r_{12}^{1-3\e},
\ee
where the numerical factor (after the convenient factoring of $\Omega_d$, and some simplification) is found to be
\begin{align}
N^{(d)}_{u_1u_2  g}&= \frac{d-2}{4(d-4)^3}\bigg[ -\frac{2}{d-3}
\nonumber\\[1ex]
&\quad + 2 \pi \cot\frac{d\pi}{2} \frac{\Gamma(\frac{3d}{2}-5)}{\Gamma(\frac{d}{2}-2)\Gamma(d-2)}\bigg].
\end{align}
The $\e$-expansion of the latter numerical factor is
\be
N^{(d)}_{u_1u_2  g}=\frac1{2\e}+2 +\frac14 (18+\pi^2) \e + O(\e^2).
\ee

Having the analytical expressions of $ I^{(d)}_{u_1u_2 g}$ and  $I^{(d)}_{u_1u_2}$, the formula \eqref{g2reduction}
reduces the computation of $ I^{(d)}_{u_1u_2 g^2}$ to that of $I^{(d)}_{u_1u_2 {\bar g}^2}$. Though ${\bar g}_d $
has a finite limit when $d=3+\e\to3$, and the coefficient of $I^{(d)}_{u_1u_2 {\bar g}^2}$ is finite as $\e\to 0$ [so that it is enough
to control $I^{(d)}_{u_1u_2 {\bar g}^2}$ to $O(\e^1)$ to get $I^{(d)}_{u_1u_2 { g}^2}$ to $O(\e^1)$], there are subtleties linked
to the {\it non uniformity} of $\lim_{\e \to 0} {\bar g}_d $. Indeed, one must treat separately the contributions to the spatial integral
$I^{(d)}_{u_1u_2 {\bar g}^2}$ coming from some (large but) finite ball, say $| {\bm x}| < R$, and the contribution from spatial infinity, i.e.\ for  $| {\bm x}| > R$.
(Henceforth, we take the origin of space at the midpoint between  ${\bm x}_1$ and ${\bm x}_2$, because this
significantly simplifies the asymptotic analysis at spatial infinity.)
More precisely, let us write [where the factor  $(1-\e)^2/\Omega_d$ is added for convenience]
\be
\frac{(1-\e)^2}{\Omega_d} I^{(d)}_{u_1u_2 {\bar g}^2}
= \frac{(1-\e)^2}{\Omega_d} I^{(d) \, < }_{u_1u_2 {\bar g}^2} + \frac{(1-\e)^2}{\Omega_d} I^{(d) \, >}_{u_1u_2 {\bar g}^2},
\ee
where
\be
\frac{(1-\e)^2}{\Omega_d} I^{(d) \, < }_{u_1u_2 {\bar g}^2}
= \frac{1}{\Omega_d} \int_{| {\bm x}| < R} \md^d x \, u_1 \, u_2 \, \left( (1-\e) {\bar g}_d  \right)^2
\ee
and
\be
\frac{(1-\e)^2}{\Omega_d} I^{(d) \, > }_{u_1u_2 {\bar g}^2}
= \frac{1}{\Omega_d} \int_{| {\bm x}| > R} \md^d x \, u_1 \, u_2 \, \left( (1-\e) {\bar g}_d \right)^2.
\ee
The first ($<$) integral has a limit as $\e \to 0$ which is simply given by
\be \label{id<}
\lim_{\e \to 0} \frac{(1-\e)^2}{\Omega_d} I^{(d) \, < }_{u_1u_2 {\bar g}^2}
= \frac1{4\pi} \int_{| {\bm x}| < R} \md^3x \, r_1^{-1} r_2^{-1} \left(\ln\frac{s}{2}\right)^2.
\ee
To compute the $\e \to 0$ limit of the second ($>$) integral, we need an approximation to $(1-\e) {\bar g}_d$ that is valid near
spatial infinity, and in $d$ (rather than $3$) dimensions. For orientation, we recall that in $d=3$, the explicit knowledge of
$\lim_{\e \to 0} {\bar g}_d \equiv  {\bar g}_3= \ln\frac{s}{2}$ allows one to compute  (when $r \equiv | {\bm x}|  \to \infty$)
\be
{\bar g}_3 = \ln \left(r+ \frac{r_{12}}{2} + O\left(\frac1{r}\right) \right) = \ln r +  \frac{r_{12}}{2 r} + O\left(\frac1{r^2}\right).
\ee
The $d$-dimensional analog of this asymptotic expansion  is obtained by combining the term-by-term inverse 
Laplacian of the asymptotic expansion of the source of $g_d$, namely
\be
\Delta { g}_d =  r_1^{2-d} \,  r_2^{2-d} = r^{4-2 d} \left(1 + O\left(\frac1{r^2}\right)  \right),
\ee
with the general multipolar-expansion formula for the ($d$-dimensional) Poisson integral of an extended
(but fast-decreasing at spatial infinity) source $s( {\bm x})$:
\begin{align}
\left[ \Delta^{-1} s  \right] ( {\bm x}) &= - \frac{{\tilde k}}{4 \pi} \int \md^d y \, |{\bm x} - {\bm y}|^{2-d} s( {\bm y})
\nonumber\\
&\approx  - \frac{{\tilde k}}{4 \pi} \frac{ \int \md^d y \,  s( {\bm y})}{  r^{d-2}} \left(1 + O\left(\frac1{r}\right)  \right).
\end{align}
Actually, as the relevant source, $r_1^{2-d} \,  r_2^{2-d}$, is not fast-decreasing when $d \approx 3$, one needs to adequately
combine the two informations.\footnote{This way of combining two expansions to get the proper behavior
of $d$-dimensional inverse Laplacians near spatial infinity was devised by Gerhard Sch\"afer and one of us (PJ)
and it was never used so far in a published work.
It is an IR analogue of the $d$-dimensional UV local analysis introduced in Ref.\ \cite{Damour:2001bu}
and completed (by the use of an explicit expression for the homogeneous contributions) in Appendix C4 of Ref.\ \cite{Jaranowski:2015lha}.}
This leads to
\begin{align}
\Delta^{-1} \left( r_1^{2-d} \,  r_2^{2-d} \right)&= \frac{ r^{6-2d} }{(6-2d)(4-d)}
\nonumber \\[1ex]
&\quad + \frac{r_{12}^{4-d}}{2(4-d)} r^{2-d} + \cdots,
\end{align}
which is equivalent to
\be
(1-\e) {\bar g}_d = \frac{1-r^{-2 \e}}{2 \e}+ \frac{r_{12}^{1-\e}}{2} r^{-1-\e} + \cdots.
\ee
Inserting the latter asymptotic expansion [together with the $(2-d)$th power of $r_1r_2=r^2(1+O(1/r^2)) $,
and $\md^d x= \Omega_d r^{d-1}\md r$]
within the definition of $\frac{(1-\e)^2}{\Omega_d} I^{(d) \, > }_{u_1u_2 {\bar g}^2}$ 
allows one to estimate the latter integral by means of a computable radial integral which yields
\begin{align}
\frac{(1-\e)^2}{\Omega_d} I^{(d) \, > }_{u_1u_2 {\bar g}^2}
&= \frac{r_{12}^{1-\e} }{8 \, \e^2} - f\left(R+\frac{ r_{12}}{2}\right)
\nonumber\\[1ex]
&\quad+O\left(\frac1R\right)+ O(\e),
\end{align}
where we introduced the function (of one variable)
\be
f(r) \equiv r \left( \ln^2 r - 2 \ln r + 2  \right).
\ee
The appearance of the term $- f(R+\frac{ r_{12}}{2})$ is exactly what is needed to define the Hadamard-regularization
of the usual $3$-dimensional integral \eqref{id<}. Indeed, one checks that the difference
\be \label{hadamard}
\frac1{4\pi} \int_{| {\bm x}| < R} \md^3 x \, r_1^{-1} r_2^{-1} \left( \ln\frac{s}{2}\right)^2 - f\left(R+\frac{ r_{12}}{2}\right)
\ee
has a finite limit as $R \to \infty$. 

The next step is to recognize that the limit as $R \to \infty$ of \eqref{hadamard} can be alternatively defined by an analytic
continuation as $ c \leadsto 0$ of the integral (over the full 3-dimensional space) of
\be \label{diffgRf}
\frac1{4\pi} \int \md^3 x \, r_1^{-1} r_2^{-1} \left(\frac{s}{2}\right)^c \left( \ln\frac{s}{2}\right)^2.
\ee
A subtle point here is that one obtains such a simple result [with the one-scale counterterm $f(R+\frac{ r_{12}}{2})$] only
when the exponents of $r_1$ and $r_2$ are both equal to $-1$. (Indeed, this guarantees that asymptotically
$\frac{\md^3 x}{4\pi} \, r_1^{-1}r_2^{-1} = \md r = \md S$, with $S \equiv r +\frac{ r_{12}}{2}$.)

Let us then consider the following version of the generalized Riesz formula
(with a normalization which is convenient for our present purpose)
\begin{align}
{\widehat I}[a,b,c] &\equiv \frac1{4\pi} \int  \md^3 x \, r_1^{a} \, r_2^{b} \, \left( \frac{s}{2} \right)^c
\nonumber \\[1ex]
&= {\widehat R}[a,b,c] \,   r_{12}^{3+a+b+c}.
\end{align}
The restriction of the generalized Riesz formula to the special case $a=b=-1$ (keeping $c$ away from zero) then  yields
the following very simple result\footnote{The simplicity of this result allows us to expand in powers of $c$ (i.e.\ to compute
and integral involving integer powers of $\ln s$ by elementary means). The expansion in more general cases where $(a, b)$
 deviate from $(-1,-1)$ (or other integer pairs) by $O(c)$ can also be analytically performed, though via
 more sophisticated techniques \cite{Moch:2001zr,Huber:2007dx}.}
\be
{\widehat I}[-1,-1,c]=  - \frac{  r_{12}^{1+c} }{1+c}.
\ee
The latter result can be easily derived from scratch by using elliptic coordinates. Indeed, in elliptic coordinates
($\xi\equiv\frac{r_2+r_1}{r_{12}}$, $\eta\equiv\frac{r_2-r_1}{r_{12}}$)
one has $\md^3x/(r_1 r_2)= \frac{r_{12}}{2}\,\md\xi \,\md\eta \,\md\phi$.
One then deduces that
\be
\left[ \frac{ \D^2 {\widehat I}[-1,-1,c]}{\D c^2} \right]_{c \leadsto 0} = - f(r_{12})
\ee
or, equivalently, in view of the previous reasonings, that 
\bea \label{valuehadamard}
&& \lim_{R\to \infty} \left[\frac1{4\pi} \int_{| {\bm x}| < R}\md^3 x \, r_1^{-1}r_2^{-1} \left( \ln\frac{s}{2}\right)^2 - f\left(R+\frac{ r_{12}}{2}\right) \right]
\nonumber \\[1ex]
&& \quad  =- f(r_{12}).
\eea
Actually, the latter result can also be more directly derived simply by evaluating the $r<R$-truncated generalized
Riesz integral in elliptic coordinates, which yields [for large $R$, modulo $O(1/R)$]
\begin{align}
&\frac1{4\pi} \int_{| {\bm x}| < R}  \md^3 x \, r_1^{-1} r_2^{-1} \left( \frac{s}{2} \right)^c
\nonumber\\[1ex]
&\quad = \frac1{1+c} \left[ \left(R+\frac{r_{12}}{2}\right)^{1+c} - r_{12}^{1+c} \right].
\end{align}
Differentiating  this result twice with respect to $c$ then yields \eqref{valuehadamard}.

Finally, putting together our results we can analytically compute the first three terms of the $\e$ expansion of 
the ${\bm x}$-space integral $I^{(d)}_{u_1u_2 { g}^2}$, namely
\be
\frac1{\Omega_d} I^{(d)}_{u_1u_2  g^2}({\bm x}_1 - {\bm x}_2) = N^{(d)}_{u_1u_2  g^2} \ r_{12}^{1-5\e},
\ee
where the numerical factor (after the convenient factoring of $\Omega_d$) is found to be
\be
N^{(d)}_{u_1u_2  g^2}= -\frac14 \left[ \frac1{\e^2} + \frac{7}{\e} + 30 + \pi^2 + O(\e) \right].
\ee
The Fourier transform (with respect to ${\bm x}_1 - {\bm x}_2$) of this ${\bm x}$-space integral, and the addition of the
various needed conventional, normalization coefficients then yields the first three terms of the $\e$ expansion of 
the master integral $\cM_{3,6}$, namely
\be
\cM_{3,6}({\bm p})=  \widehat{\cM}_{3,6}  \ |{\bm p}|^{4 \e -4},
\ee
where (with $\gamma$ denoting Euler's constant)
\be
 \widehat{\cM}_{3,6} =(4\pi)^{-4-2 \e} \frac{e^{2 \gamma \, \e}}{2} \left[\frac1{\e^2} - \frac1{\e} + \frac{\pi^2}{12}-8 + O(\e)   \right].
\ee
Our reasoning has analytically proven the latter expansion (which agrees with the result of  \cite{Foffa:2016rgu}), and has, actually, reduced it to the evaluation of more
elementary integrals: notably the two-loop integral $I^{(d)}_{u_1u_2  g}$, and the integrals involving $\ln^2 (s/2)$ discussed
above, which were, actually, reduced to trivial integrals when using elliptic coordinates (and these trivial integrals did not
involve any irrational coefficients).

Separately from the technical issue of analytically evaluating such integrals, let us note again that the evaluation of the
contentious contributions to the four-loop effective action discussed in the previous sections involved only IR convergent integrals,
while the master integral $\cM_{3,6}$ is IR divergent (as shows up in its singular behavior as $\e \to 0$). This indicates
that choosing $\cM_{3,6}$ as one of the basis of elementary master integrals is probably not an optimal choice.

\section{Conclusions}

We have shown that  remarkable cancellations take place within the  four-loop, 4PN, $O(G^5)$
``static''\footnote{In the sense of being independent both on velocities and their
time derivatives.}  contribution to the original, higher-time-derivative, 
harmonic-coordinates effective action of a gravitationally interacting binary point-mass system.
Namely, the subset of diagrams $\propto G^5 m_1^3 m_2^3/(c^8 r_{12}^5)$ (denoted $L_{33}$, $L_{49}$, $L_{50}$ 
in Ref.\ \cite{Foffa:2016rgu}) that individually involve transcendental coefficients
$ \in \mathbb{Q}[\pi^2] =\mathbb{Q}[\zeta(2)]$ cancell against each other to leave a final, rational coefficient $+ \frac{40}{3}$.
On the one hand, this finding corrects a recent claim of Ref.\ \cite{Foffa:2016rgu}, which found a final coefficient for the
same terms equal to $\frac{1112}{9} - \frac{32}{3} \pi^2$. On the other hand, it confirms a previous lower-order finding
of \cite{Foffa:2011ub}, namely the fact that the corresponding highest-power-of-$G$, static terms at the previous PN level
[three-loop, 3PN, $O(G^4)$ level] did not involve any $\pi^2$ dependence, by contrast with the two-loop, 3PN, $O(G^3 v^2)$ terms.
We leave to future work a deeper understanding of the rational-coefficient nature of such, highest-$G$-order,
static terms at each PN order. As pointed out by Foffa \textit{et al.} \cite{Foffa:2011ub,Foffa:2016rgu}, the same terms (at 3PN and 4PN)
happen to be finite at $d=3$ (in dimensional regularization). At 4PN, this finiteness comes after the cancellation of
poles $\propto 1/(d-3)$ present in individual diagrams. The latter cancellations can be seen rather easily, at 4PN, from the
explicit ${\bm x}$-space  expressions that we have given above for all the static 4PN diagrams (and not only $L_{33}$, $L_{49}$, $L_{50}$).

The cancellations discussed above are specific to the harmonic-gauge computation of the effective action. E.g. the situation is
different in ADM gauge, where there are static, three-loop, 3PN, $O(G^4)$ terms involving $\pi^2$, as well as
static, four-loop, 4PN, $O(G^5)$ terms involving $\pi^2$. It remains, however, true that the effective action
for the gravitational interaction of point masses exhibit a remarkably small level of transcendentality.
At one and two loops (at 1PN and 2PN), the action involves only rational coefficients. The 3PN, three-loop
level introduces $\mathbb{Q}[\zeta(2)]$ coefficients, and this transcendentality level does not increase
when going to the 3PN, four-loop level. Very-high-PN-order, analytical gravitational self-force studies of the EOB
Hamiltonian \cite{Bini:2014nfa,Bini:2015bla,Kavanagh:2015lva} have shown (for a subset of the diagrams)
that the transcendentality level increases only quite slowly as the loop number (equal to the PN level) increases:
the $\mathbb{Q}[\zeta(4)] = \mathbb{Q}[\pi^4]$
level is reached at six loops, and $\zeta(3)$ first appears at the seven-loop order.
[Here, we are (roughly) subtracting the effects linked to non-local-in-time interactions
which introduce Euler's constant $\gamma$ and logarithms.]
We leave to future work a better understanding of such facts.

Separately from the interest of finding special structures hidden in the gravitational
effective action, our work  provides a confirmation of the correctness of the 4PN-level 
$O(m_1^3 m_2^3)$ sector of the harmonic-coordinates action of \cite{Bernard:2015njp}.
This confirmation is independent of that following from its previously checked
agreement with the corresponding sector of the 4PN, ADM action of \cite{Damour:2014jta,Jaranowski:2015lha}. 
[In terms of the $\mu$-reduced Hamiltonian, this corresponds to $ O(\nu^2)$ terms that had been first derived  in Ref.\ \cite{Jaranowski:2013lca}.]
Having such independent confirmations is always useful.
It would be useful that a full, independent 4PN, EFT-based computation of the 4PN effective action be performed.
However, in view of the complications (and sign  dangers) brought by working with purely imaginary propagators, and
corresponding $i$-decorated vertices, we would advocate (as explained at the end of  Sec.\ II above) to work
with {\it real} propagators $\cG = - {\cal K}^{-1}$, and corresponding $i$-free vertices (when viewed in ${\bm x}$-space).

Let us finally comment on the technicalities of the explicit, 4PN computation. We have shown in Sec.\ V above that the four-loop
master integral $\cM_{3.6}$ selected as basis element in \cite{Foffa:2016rgu}, and that could only be numerically computed
in the latter reference, could be analytically computed by means of what has been the standard tool
in ADM computations since the 3PN level, namely the generalized Riesz formula \cite{Jaranowski:1997ky}. It is remarkable that
a tool set up for the three-loop level can analytically deal with a four-loop integral that resisted the state-of-the-art
technologies in multi-loop computations. We think that this is due to two main facts (besides the special structure of
the gravitational vertices): (i) our use of ${\bm x}$-space integration\footnote{We note in this respect that
the first five four-loop master integrals in Fig.\ 3 of \cite{Foffa:2016rgu} are trivial to compute in ${\bm x}$-space.
Indeed, repeated lines between two points just mean a power of $r_{12}$, $r_1$ or $r_2$, without any needed integration.
The only needed integrations in ${\bm x}$-space go with the number of intermediate vertices;
for instance,  $\cM_{1,1}$, $\cM_{1,2}$, $\cM_{1,3}$ and $\cM_{1,4}$ only involve one intermediate-point integration, so that they are immediately derived from the normal
Riesz formula, while $\cM_{3,6}$
involves three intermediate-points integrations.} and (ii) the fact that the repeated differentiation of the
generalized Riesz formula with respect to the power of $s\equiv r_1+r_2+r_{12}$ allows one to compute integrals
that can show up at an arbitrary high loop order. Indeed, the $n$th derivative with respect to the power of $s$ generates
(3-dimensional) integrals of the type, say
\bea
I_{n_1,n_2,n} &\sim&  \int d^3 x \,  r_1^{-n_1} r_2^{-n_2} (\ln s)^n
\nonumber\\[1ex]
&\subset&  r_1^{-n_1} r_2^{-n_2} \left( \Delta^{-1} r_1^{-1} r_2^{-1} \right)^n.
\eea
When $n_1=n_2=1$ and $n=2$ this corresponds to the four-loop $\cM_{3.6}$ diagram. When taking higher values of $n$,
$I_{n_1,n_2,n}$ describes higher-loop master integrals.

\begin{acknowledgments}

T.D. thank Pierre Vanhove for informative discussions, and useful references, on Feynman integrals.
We thank Stefano Foffa for clarifying the precise meaning of the notation for the kinetic terms in the bulk action for $\sigma_{ij}$ and $\phi$,
and notably $(\overrightarrow{\nabla}\sigma)^2$.
The work of P.J.\ was supported in part by the Polish NCN Grant No.\ UMO-2014/14/M/ST9/00707.

\end{acknowledgments}

\appendix

\section{Some useful formulas}
\label{appendixA}

\subsection{$d$-dimensional results}

The area of the $(d-1)$-dimensional unit sphere in $\mathbb{R}^d$ reads
\be
\Omega_d= \frac{2 \pi^{d/2}}{\Gamma(d/2)}.
\ee
It is convenient to introduce the constant
\be
{\tilde k} \equiv  \frac{\Gamma(\frac{d}{2}-1)}{\pi^{\frac{d}{2}-1}},
\ee
such that
\be
{\tilde k} \, \Omega_d = \frac{4 \pi}{d-2}.
\ee
Then the $d$-dimensional Newtonian potential $u_a \equiv   r_a^{2-d}$ fulfills the equation
\be
\Delta \left( {\tilde k} \, r_a^{2-d} \right) = - 4\pi \delta_a.
\ee

The (ordinary) Riesz formula in $d$ dimensions reads
\begin{subequations}
\label{riesz}
\be
\int \md^dx \, r_1^a r_2^b= \pi^{\frac{d}{2}} \Gamma^{(6)}_{a,b}(d) r_{12}^{a+b+d},
\ee
with
\be
\Gamma^{(6)}_{a,b}(d) \equiv \frac{\Gamma\left(\frac{a+d}{2}\right)
\Gamma\left(\frac{b+d}{2}\right)
\Gamma\left(-\frac{a+b+d}{2}\right)}
{\Gamma\left(-\frac{a}{2}\right)
\Gamma\left(-\frac{b}{2}\right)
\Gamma\left(\frac{a+b+2d}{2}\right)}.
\ee
\end{subequations}
A three-dimensional generalization of the Riesz formula \eqref{riesz} for integrands of the form
$r_1^a r_2^b (r_1+r_2+r_{12})^c$ was derived in Ref.\ \cite{Jaranowski:1997ky}.
It reads
\begin{subequations}
\be
\int \md^3x \, r_1^a r_2^b (r_1+r_2+r_{12})^c
= 2\pi R(a,b,c)\,r_{12}^{a+b+c+3},
\ee
where
\begin{align}
R(a,b,c) &\equiv
\frac{\Gamma\left(a+2\right)\Gamma\left(b+2\right)
\Gamma\left(-a-b-c-4\right)}{\Gamma\left(-c\right)}
\nonumber\\[1ex]
&\quad
\times \Big[ I_{1/2}\left(a+2,-a-c-2\right)
\nonumber\\[1ex]
&\quad + I_{1/2}\left(b+2,-b-c-2\right)
\nonumber\\[1ex]
&\quad - I_{1/2}\left(a+b+4,-a-b-c-4\right) - 1 \Big].
\end{align}
\end{subequations}
The function $I_{1/2}$ is defined as follows:
\be
I_{1/2}\left(x,y\right) \equiv \frac{B_{1/2}\left(x,y\right)}{B\left(x,y\right)},
\ee 
where $B$ is the Euler beta function and $B_{1/2}$ is the incomplete beta function
which can be expressed in terms of the Gauss hypergeometric function $_2F_1$:
\be
B_{1/2}\left(x,y\right)=\frac{1}{2^x x}\,
{_2F_1}\!\!\left(1-y,x;x+1;\frac{1}{2}\right).
\ee

The $d$-dimensional Fourier transform of a power reads:
\be
\int   {\bar \md} p \, e^{i {\bm p} \cdot {\bm r}} \frac{\Gamma(a)}{(p^2)^a}
= \frac1{\pi^{\frac{d}{2}} 2^{2a}} \frac{\Gamma(\frac{d}{2}-a)}{(r^2)^{\frac{d}{2}-a}}.
\ee
The result of  \cite{Chetyrkin:1980pr} (and \cite{Suzuki:2014hda}) for the ${\bm p}$-space version
of the two-loop diagram of Fig. \ref{fig:u1u2g} reads
\be
\int \frac{{\bar \md}k\,{\bar \md}\ell} {k^2 \ell^2 (k-p)^2 (\ell-p)^2 (k-\ell)^2}
= \frac{ (p^2)^{d-5} }{(4\pi)^d} \Gamma^{(8)}(d),
\ee
where
\begin{align}
&\Gamma^{(8)}(d)
\equiv \frac{\Gamma(\frac{d}{2} - 2)^2 \Gamma(2 - \frac{d}{2}) \Gamma(\frac{d}{2} - 1)}{\Gamma(d - 2)}
\nonumber\\[1ex]
&\times \left(\frac{\Gamma(3 - \frac{d}{2})\Gamma(\frac{d}{2} - 1)}{\Gamma(d - 3)}
- \frac{\Gamma(d - 3) \Gamma(5 - d)}{\Gamma(3 - \frac{d}{2}) \Gamma(\frac{3d}{2} - 5)} \right).
\end{align}

\subsection{Distributions in $d=3$ dimensions}

In Sec.\ IV we have to compute different distributional derivatives.
We collect here formulae which can be used for this goal.
Let us start from identities involving Dirac delta distrubution and its derivatives:
\begin{subequations}
\begin{align}
f \,\delta_a &= f_\text{reg}(\bm{x}_a)\, \delta_a,
\\[1ex]
f \,\nabla\delta_a &= -(\nabla f)_\text{reg}(\bm{x}_a)\, \delta_a
+ f_\text{reg}(\bm{x}_a)\, \nabla\delta_a,
\\[1ex]
f \,\Delta\delta_a &= (\Delta f)_\text{reg}(\bm{x}_a)\, \delta_a
- 2 (\nabla f)_\text{reg}(\bm{x}_a)\cdot\nabla\delta_a
\nonumber\\[1ex]
&\qquad + f_\text{reg}(\bm{x}_a)\, \Delta\delta_a.
\end{align}
\end{subequations}
Because usually the function $f$ for which the above identities are used
is singular at $\bm{x}=\bm{x}_a$, the symbol $f_\text{reg}(\bm{x}_a)$
means the regularized ``partie finie'' value of the function $f$ at $\bm{x}=\bm{x}_a$
(for its definition and properties see, e.g., Appendix A4 of Ref.\ \cite{Jaranowski:2015lha}).

We have also employed distributional derivatives
to calculate first and second partial derivatives of homogeneous functions $1/r_a$, $1/r_a^2$, and $1/r_a^3$
(for derivation and properties see, e.g., Appendix A5 of Ref.\ \cite{Jaranowski:2015lha}).
The first partial derivatives read
\begin{subequations}
\begin{align}
\label{e41a1}
\partial_i\frac{1}{r_a} &= -\frac{n_a^i}{r_a^2},
\\[1ex]
\label{e41a2}
\partial_i\frac{1}{r_a^2} &= -\frac{2n_a^i}{r_a^3},
\\[1ex]
\label{e41a3}
\partial_i\frac{1}{r_a^3} &= -\frac{3n_a^i}{r_a^4} - \frac{4\pi}{3}\partial_i\delta_a.
\end{align}
\end{subequations}
The second partial derivatives are
\begin{subequations}
\begin{align}
\label{e41b1}
\partial_i\partial_j\frac{1}{r_a} &= \frac{3n_a^in_a^j-\delta^{ij}}{r_a^3} - \frac{4\pi}{3}\delta_{ij}\delta_a,
\\[1ex]
\label{e41b2}
\partial_i\partial_j\frac{1}{r_a^2} &= \frac{2(4n_a^in_a^j-\delta^{ij})}{r_a^4},
\\[1ex]
\label{e41b3}
\partial_i\partial_j\frac{1}{r_a^3} &= \frac{3(5n_a^in_a^j-\delta^{ij})}{r_a^5}
- \frac{2\pi}{15} \left( 16\partial_i\partial_j\delta_a + 3\delta_{ij}\Delta\delta_a \right).
\end{align}
\end{subequations}
Tracing the above formulas yields
\begin{subequations}
\begin{align}
\label{e41c1}
\Delta\frac{1}{r_a} &= -4\pi\delta_a,
\\[1ex]
\label{e41c2}
\Delta\frac{1}{r_a^2} &= \frac{2}{r_a^4},
\\[1ex]
\label{e41c3}
\Delta\frac{1}{r_a^3} &= \frac{6}{r_a^5}
- \frac{10\pi}{3}\Delta\delta_a.
\end{align}
\end{subequations}

As an application of the above formulas let us note a useful expression which shows how to compute
the Laplacian of the product of a (singular at $\bm{x}=\bm{x}_a$) function $f$ and $1/r_a^3$:
\begin{align}
\Delta\left(f\frac{1}{r_a^3}\right) &= \Delta\left(f\frac{1}{r_a^3}\right)\bigg|_\textrm{ord}
- \frac{2\pi}{3}(\Delta f)_\text{reg}(\bm{x}_a)\, \delta_a
\nonumber\\[1ex]
&\quad + 4\pi(\nabla f)_\text{reg}(\bm{x}_a)\cdot\nabla\delta_a
- \frac{10\pi}{3}  f_\text{reg}(\bm{x}_a)\, \Delta\delta_a,
\end{align}
where $\Delta\left(f/r_a^3)\right)|_\textrm{ord}$ means the Laplacian computed using
standard (i.e.\ non-distributional) rules of differentiations.

\end{document}